\documentclass[11pt]{article}
\usepackage[left=3cm, right=3cm, top=3cm, bottom=3cm]{geometry}
\usepackage[utf8]{inputenc}
\usepackage[T1]{fontenc}
\usepackage{lmodern}
\usepackage{amssymb,siunitx,amsmath,babel,graphicx,tabularx,hyperref,multicol,hhline,multirow,titlesec,titling,mathtools,nccmath,setspace,ragged2e,caption,subcaption,float}
\usepackage{booktabs}
\usepackage[sort&compress]{natbib}
\usepackage[inkscapelatex=false]{svg}
\DeclareMathOperator*{\argmax}{arg\,max}
\usepackage{threeparttable}
\titlespacing{\section}{2pt}{1\parskip}{0\parskip}
\titlespacing{\subsection}{2pt}{.5\parskip}{-.75\parskip}
\titlespacing{\subsubsection}{2pt}{.25\parskip}{-.75\parskip}
\makeatletter
\let\@msm@th@eqref\eqref
\renewcommand{\eqref}[1]{
  \begingroup
  \leavevmode
  \color{blue}
  \hypersetup{linkbordercolor=[named]{blue}}
  \@msm@th@eqref{#1}
  \endgroup
}
\makeatother

\newcommand{\acknowledgements}[1]{
    \textbf{Acknowledgement: }
  \small#1
}

\newcommand\ddfrac[2]{\frac{\displaystyle #1}{\displaystyle #2}}
\newcommand{\subtitle}[1]{
  \posttitle{
    \par\end{center}
    \begin{center}\large#1\end{center}
    \vskip0.3em}
}

\usepackage{chngcntr} 
\counterwithout{table}{section} 
\usepackage{apptools} 
\AtAppendix{
    \setcounter{table}{0} 
}

\title{\textbf{Evaluating the Public Pay Gap:} A Comparison of Public and Private Sector Wages in France}
\author{Riddhi Kalsi \thanks{Sciences Po, Paris; \texttt{riddhi.kalsi@sciencespo.fr}}}
\date{\vspace{-10ex}}
\usepackage{titling}
\newcommand\blfootnote[1]{
  \begingroup
  \renewcommand\thefootnote{}\footnote{#1}
  \addtocounter{footnote}{-1}
  \endgroup
}

\begin{document} 
\maketitle

\begin{abstract}
    \vspace{-1cm}
   \setstretch{1.25}
      \noindent  \justify{This paper resolves the empirical puzzle in the public–private wage literature: why studies using similar data reach contradictory conclusions about wage premiums and penalties. Utilizing rich French administrative panel data (2012–2019), this study has two main contributions: first, it presents a set of new, intuitive yet previously undocumented stylized facts about wage dynamics, sectoral mobility, and gender differences across sectors. The results reveal that the modest hourly wage gaps conceal substantial disparities in lifetime earnings and employment stability. Women, in particular, gain a significant lifetime earnings advantage in the public sector, driven by higher retention, better-compensated part-time work, and more equitable annual hours compared to the private sector, where gender gaps remain larger, especially for those with higher education. In contrast, highly educated men experience a lifetime penalty in public employment due to rigid wage structures. By flexibly modeling sectoral transitions, transitions into and out of employment, and earnings heterogeneity using an Expectation-Maximization algorithm, this study shows that both premiums and penalties depend systematically on gender, education, and labor market experience. The analysis reveals that significant unobserved heterogeneity remains in wage dynamics. These findings unify prevailing narratives by providing a comprehensive, descriptive account of sectoral differences in transitions, part-time work and wages by gender. }  
\end{abstract}
\vspace{2cm}
\blfootnote{\acknowledgements{This work was supported by the European Research Council (grant reference ERC-2020-ADG-101018130) and the Agence Nationale de la Recherche (grant reference ANR-19-CE26-0007-01) awarded to Jean-Marc Robin.}}
    \pagebreak

This empirical puzzle stems from valuable but distinct methodological approaches and sample selection criteria to studying public-private wage differentials. Most studies fall into two distinct camps: those using fixed-effects regressions on hourly wages\footnote{\citep{melly2006public} and more recent work \citep{melly2006public}} and those employing dynamic lifetime approaches on monthly or annual wages\footnote{\citep{beffy,ppgapeur}}. However, neither approach simultaneously incorporates the key elements that explain contradictory empirical findings. At its heart, the public-private wage gap question is: ``For a given person, what is the effect of working in the public sector on their wage, compared to if they had worked in the private sector?". The epistemological problem is that we can never observe this counterfactual. We cannot see the same person in both sectors at the same time. Therefore, any statistical method is an attempt to construct a \textit{plausible} counterfactual group to compare against.

The choice of method dictates the nature of that counterfactual and thus defines what we are actually claiming to know. Fixed effects regressions estimate the wage gap for people who switch sectors, averaging over switchers their wages before and after the sectoral change while controlling for innate, time-invariant traits. Lifetime earnings approaches with selection models estimate the expected gap for the entire population, using statistical techniques to simulate what a typical person would earn in either sector by accounting for job loss risk and the preexisting differences that lead people to choose one sector over the other.

The answer to this empirical conundrum thus lies in the methodology used. Fixed-effects studies on hourly wages, while controlling for unobserved heterogeneity, focus on hourly wages and miss the differential impact of hours worked—particularly relevant given women’s higher propensity for part-time work. Meanwhile, lifetime approaches account for job security through employment transitions but have either been limited to male-only samples, or have not allowed for free movement between the sectors. This study uses the former method only in section \ref{descriptives_section} and the latter for the main results due to its advantages in taking job stability and annual wages (as a result of heterogeneous labor supply) into account, and statistically modeling sectoral stayers as well as movers. It turns out, that allowing for sector switching in a lifetime earnings approach doesn't make a big difference in estimating lifetime earnings in either sector as this study shows that sector-switching is relatively rare in the French context (in subsection \ref{sebsec:switch}). A note of caution here, the counterfactual lifetime earnings presented in this analysis assume that neither public nor private sector parameters (in terms of wages, employment, job retention and mobility) change. This gives the analysis a more descriptive rather than causal flavor.

This study makes three key contributions. First, I demonstrate that seemingly modest hourly wage gaps between public and private sectors mask profound disparities in lifetime earnings, with the public sector exhibiting gender gaps 13\% lower than the private sector on average across earnings quantiles through better retention and compensation for part-time work—a mechanism overlooked in prior hourly wage comparisons. Second, the study builds on the current literature by simultaneously examining gender dynamics, unobserved heterogeneity\footnote{A critical issue in estimating the public-private wage gap is selection bias, which arises because public sector jobs often attract individuals with different preferences, skills, or motivations compared to those in the private sector. \citep{heckmansinger1984} famously introduced the selection correction technique, which allows researchers to account for such unobserved differences and has been extended and applied by \citep{beffy} in looking at the public pay gap in France. For instance, public sector workers may have a stronger preference for job security or non-wage benefits like pensions, which can complicate comparisons with private sector employees who may prioritize higher pay or career mobility. Studies that fail to account for this selection bias risk over- or under-estimating the true wage differential. \\
Based on the above analysis, one could argue that there is self-selection of women into the public sector; that is, women prefer (or "select" themselves) to be in the public sector wherever possible. Therefore, comparing these jobs held by women in the public sector to private sector employment will mechanically give a positive public premium as agents are rational. On looking more closely, one realizes that \textit{a priori} there is no reason to assume that only women self-select into the public sector. With rational agents in society, we expect that any individual is selecting the sector most suitable for them. Hence, the same argument can be made for men’s self-selection to the public sector, and to that effect we should find similar public premia by gender and earnings percentiles, which is not the case in this setting as it is higher for women extensively and intensively. This is not a complete measure of self-selection but of the different outcomes women have compared to men.
}, unrestricted sectoral mobility, and lifetime earnings. Women disproportionately sort into stable public roles due to both observable traits like education and latent factors shaping career trajectories\footnote{This is echoed in \citep{fougere2003wants,gomes_redux,gomes2025you}}. Third, the analysis reveals a critical equity-efficiency trade-off: while the public sector insulates workers from volatility, it imposes lifetime earnings penalties on highly educated workers due to rigid wage structures. The core idea of this study is to provide a nuanced understanding of public-private earnings differences; that penalties or premiums are not universal features of either sector.

By utilizing a sequential Expectation-Maximization (EM) algorithm on French panel data that include men, women, full-time and part-time work, and allows for sector-switching, I find that public sector employment offers a significant lifetime earnings premium for women but imposes penalties for highly educated men. The methodology simultaneously estimates individual income trajectories, employment dynamics, and selection into sectors, allowing for unobserved heterogeneity in earnings patterns and sector choice. Considering lifetime values accounts for not just wage levels but also the security of income over an entire career, particularly in the face of job loss risk. \citep{Garbinti2023} highlight the importance of incorporating lifetime earnings into gender pay gap analyses, showing that short-term wage differences underestimate long-term disparities. \citep{Guvenen2021} further argue that income volatility and employment uncertainty significantly shape utility-based job preferences, particularly for risk-averse individuals who may select into the public sector for its greater stability.

This work unifies two main strands of literature on France. While \citep{mellylatest, melly2006public} report hourly wage penalties for educated workers, \citep{beffy, ppgapeur} find lifetime premia using monthly earnings and accounting for selection. This paper shows that hourly wages mask profound lifetime disparities driven by part-time work and unobserved selection. France’s institutional features—centralized wage bargaining, regulated hours, and strong job protection—position it as a prototype of coordinated market economies where similar dynamics operate.

The public sector’s role as a mitigator of gender inequality is increasingly relevant as fiscal pressures and privatization debates intensify. The gender pay gap is a well-documented phenomenon, consistently showing that women earn less than men across various sectors and countries (for example, \citep{BlauKahn2017, goldin_2014, PetersenMorgan1995}, etc.). In high-income countries, the compatibility of women’s career and family goals, facilitated by family policy, cooperative fathers, favorable social norms, and flexible labor markets, has become key drivers of fertility and participation in the labor force \citep{repec:aea:aejapp:v:2:y:2010:i:3:p:228-55}. \citep{goldin_2014} argues that while gender convergence in labor market outcomes has progressed significantly, workplace flexibility remains a key obstacle, a factor that is particularly relevant in explaining the persistent gender wage gap between public and private sectors - similarly, this study echoes the disparities in compensated hours by gender. Extensive research has explored multiple dimensions of this disparity, including differences in occupation, industry, work experience, education, and discrimination. However, a crucial aspect that warrants further attention is the variation in gender pay gaps between the private and public sectors over a lifetime. This article contributes to the existing body of literature by demonstrating that the private sector imposes a larger gender pay gap when compared to the public sector when considering lifetime earnings. Prior work highlights shrinking gender wage gaps in countries with large public sectors \citep{gen_ineq_book}, but few studies examine lifetime earnings or part-time work comprehensively. This paper resolves existing contradictions by modeling annual earnings and incorporating part-time employment—critical for women, who constitute 75\% of part-time workers in France.

   \noindent This article proceeds as follows: Section \ref{data_section} describes the data and briefly, some institutional context. Section \ref{descriptives_section} provides descriptive evidence and some stylized facts that support the statiscal model that is later specified in section \ref{model_section}. Section \ref{estimation_section} presents the unobserved heterogeneity analysis along with model fit both in- and out-of-sample. Section \ref{lifetime} documents the main results. After a discussion on points for future research in section \ref{discussion_section}, section \ref{conclusion_section} concludes.

	\section{Data} \label{data_section}
    
            \subsection{Data description} 

            This study uses the DADS-EDP tous salariés panel for 2012-2019, combining the DADS "all-employees" panel with the Échantillon Démographique Permanent (EDP). The DADS panel provides comprehensive employee and job information from annual employer declarations and state payroll files. The EDP adds demographic data from civil status records and surveys for individuals born on October 1st and 4th. The datasets are matched via registration index numbers, allowing education levels to be determined from survey responses.
            
           \noindent A key development in the data is the integration of DSN (Déclaration Sociale Nominative) information starting in 2016, which streamlined social declarations for French companies \footnote{\href{INSEE - Understanding the Nominative Social Declaration (DSN) for Better Statistical Measurement}{https://www.insee.fr/en/information/4195367?sommaire=4195376}}. By 2017, the DSN became mandatory for most private sector companies, and by 2018, 99\% of private companies had adopted it. This transition expanded the dataset's scope to include both private and public sector employees, as well as employees of individual employers, enabling more comprehensive analysis of employment and wages across the French economy.

           \subsection{Sample selection and statistics}
            The panel used in this analysis consists of 276,514 individuals (training data) which is a randomly selected quarter of the larger sample consisting of 829,232 individuals (used to check for out-of-sample model fit). The latter is a random sample that is representative of the French working population. The sample size is purposely kept large to better inform the dynamic model of the heterogeneity in mobility and income over time. The total number of panel observations including spells of non-employment total 2,065,902. Individuals aged 18-60 who have completed their education prior to or during the observed time-periods are kept to avoid peculiarities (apprenticeship for young workers, and early retirement in the public sector, for example). Employment spells that total atleast 6 months or 180 days in a year are used for the analysis \footnote{This construction includes employment spells across multiple employers within a sector, allowing for a more flexible approach than typical full-year, single-employer restrictions. If total employment in a year is less than 6 months, that year is coded as a non-employment spell. See Appendix \ref{panel_constr} for further details on panel construction.}. The log of real net annual wages or "salaire net fiscal en euros constants" is used throughout. Using real net wages provides a more accurate representation of workers' purchasing power by accounting for all social contributions and eliminating inflation effects. This measure offers a standardized basis for comparison across sectors and time periods, reflecting the real economic situation of workers more effectively than gross salary figures.

            \noindent Education is coded as a categorical variable with the three levels: high (some university qualification), medium (some vocational training post high school), and low (at most a high school degree). Table \ref{tab::educ} tabulates this classification and and provides U.S. equivalents. In the sample, approximately 69\% of individuals in the panel had a low education level, 16\% medium, and 15\% high.

            \begin{table}[htbp]
                \centering
                    \footnotesize
                \caption{Descriptive Statistics by Employment Type and Sector}
                \label{tab:descriptive1}
                \begin{tabular}{l l c c c c}
                \toprule
                & & \multicolumn{2}{c}{Full-time spells (79\%)} & \multicolumn{2}{c}{Part-time spells (21\%)} \\
                \cmidrule(lr){3-4} \cmidrule(lr){5-6}
                & & Public & Private & Public & Private \\
                \midrule
                \multicolumn{2}{l}{Employment share (\%)} & 24 & 76 & 26 & 74 \\
                \midrule
                \multirow{2}{*}{Log real net annual wage} & Mean & 10.19 & 10.14 & 9.58 & 9.32 \\
                & (sd) & (0.35) & (0.49) & (0.69) & (0.87) \\
                \multirow{2}{*}{Log real net hourly wage} & Mean & 2.69 & 2.65 & 2.56 & 2.42 \\
                &(sd)  & (0.32) & (0.42) & (0.32) & (0.41) \\
                \midrule
                \multicolumn{6}{l}{\textit{Demographics}} \\
                \multicolumn{2}{l}{Share of Women (\%)} & 61 & 36 & 84 & 77 \\
                \multicolumn{2}{l}{ Low Education (\%)} & 52 & 68 & 56 & 74 \\
                \multicolumn{2}{l}{ Med Education (\%)} & 21 & 17 & 23 & 14 \\
                \multicolumn{2}{l}{ High Education (\%)} & 28 & 15 & 21 & 12 \\
                \multicolumn{2}{l}{Ages $\leq$ 30 (\%)} & 11 & 18 & 11 & 16 \\
                \multicolumn{2}{l}{Ages 31-45 (\%)} & 41 & 43 & 44 & 39 \\
                \multicolumn{2}{l}{Ages 45+ (\%)} & 49 & 39 & 44 & 45 \\
                \midrule
                \multicolumn{2}{l}{\# panel observations} & \multicolumn{2}{c}{1,142,309} & \multicolumn{2}{c}{299,721} \\
                \multicolumn{2}{l}{\# individuals} & \multicolumn{4}{c}{276,514} \\
                \bottomrule
                \end{tabular}
            \end{table}

\noindent Table \ref{tab:descriptive1} presents descriptive statistics for full-time and part-time employment spells, broken down by public and private sectors. Full-time employment accounts for 79\% of spells, with a larger share in the private sector (76\%) compared to the public sector (24\%). Part-time employment constitutes 21\% of spells, with similar sectoral proportions (26\% public, 74\% private). The mean log real net annual wage is slightly higher in the public sector than in private. This pattern persists for part-time workers, with mean log annual wages of 9.58 in the public sector and 9.32 in the private sector. Similarly, hourly wages are marginally higher in the public sector for both full-time and part-time workers.

\noindent Demographic differences are notable. Women dominate public sector employment (61\% for full-time and 84\% for part-time) compared to the private sector (36\% and 77\%, respectively). Public sector employees also exhibit higher levels of education, with 28\% of full-time and 21\% of part-time workers having high education, versus 15\% and 12\% in the private sector. Private sector workers tend to be younger, with 18\% under 30 in full-time roles, compared to 11\% in the public sector.

\noindent Women dominate part-time employment, accounting for 79\% of part-time spells across both public and private sectors, which represents 17\% of aggregate employment during the observed period. Among women's employment spells, part-time work varies inversely with education level: 37\% for low-educated, 28\% for medium-educated, and 24\% for highly educated women. This pattern suggests an inverse relationship between educational attainment and the propensity for part-time work among women.

	\section{Stylized facts}\label{descriptives_section}	
        In these sub-sections, I present some novel evidence from the panel that motivates the estimation strategy presented in the next section. These nuances are key to understanding public-private differences and are seldom consolidated in a single narrative. The findings are intuitive, but are nonetheless important to quantify.

                \subsection{Sector-switching is rare...but not random}\label{sebsec:switch}
\begin{figure}[ht]
    \centering
    \begin{subfigure}[b]{0.45\textwidth}
        \includegraphics[width=\textwidth]{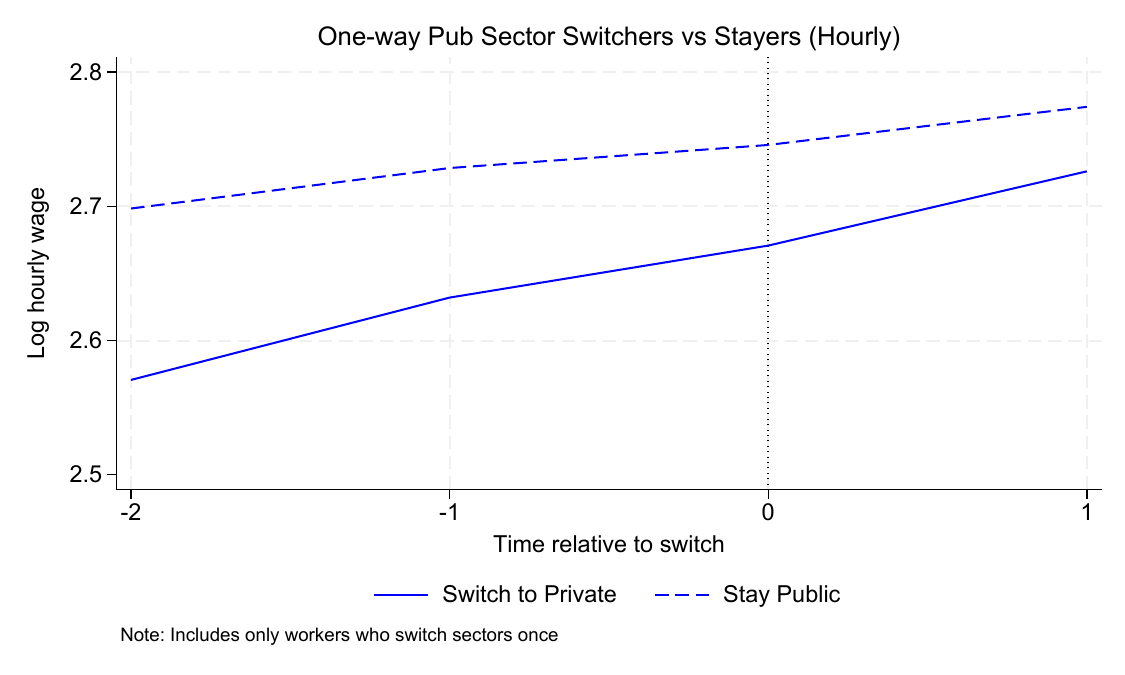}
        \caption{Hourly Wage Changes - Public Sector Switchers and Stayers}
        \label{fig:hourly_pub}
    \end{subfigure}
     \begin{subfigure}[b]{0.45\textwidth}
        \includegraphics[width=\textwidth]{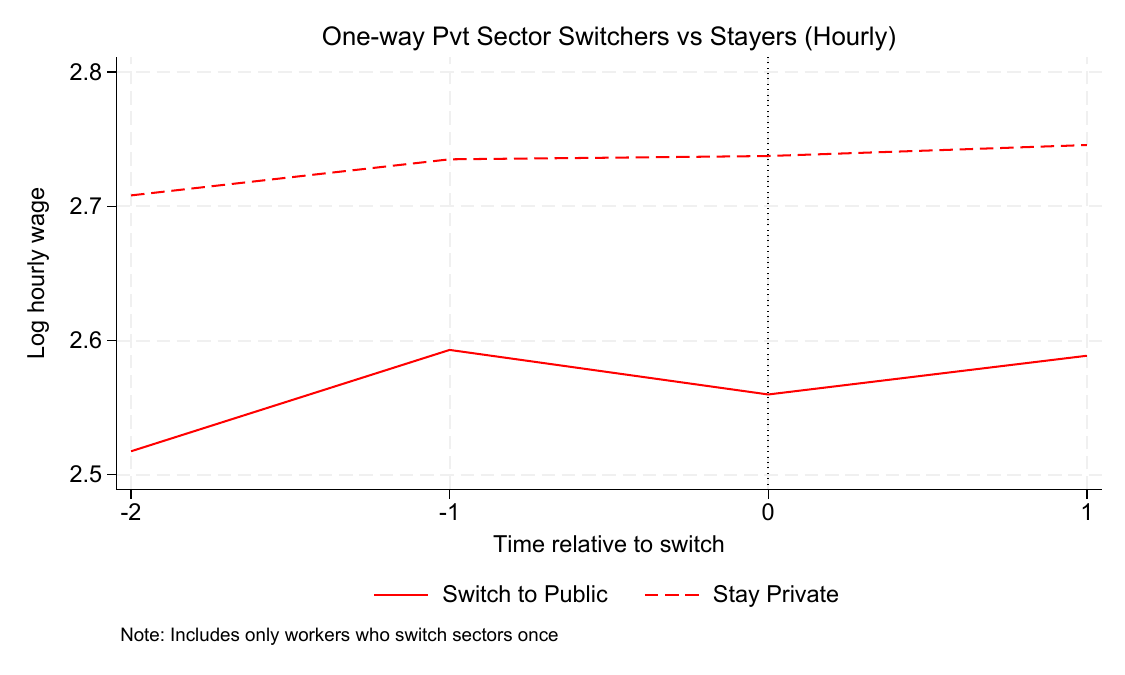}
        \caption{Hourly Wage Changes - Private Sector Switchers and Stayers}
         \label{fig:hourly_pvt}
    \end{subfigure}
    \hfill
    \begin{subfigure}[b]{0.45\textwidth}
        \includegraphics[width=\textwidth]{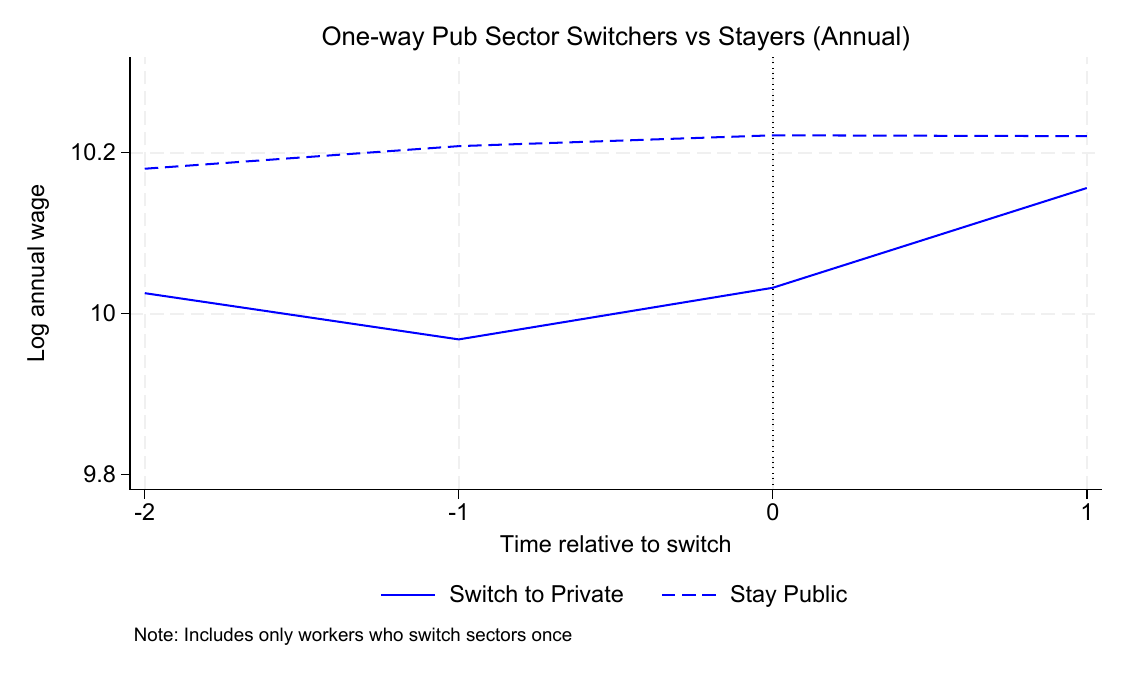}
       \caption{Annual Wage Changes - Public Sector Switchers and Stayers}
        \label{fig:annual_pub}
    \end{subfigure}
     \begin{subfigure}[b]{0.45\textwidth}
        \includegraphics[width=\textwidth]{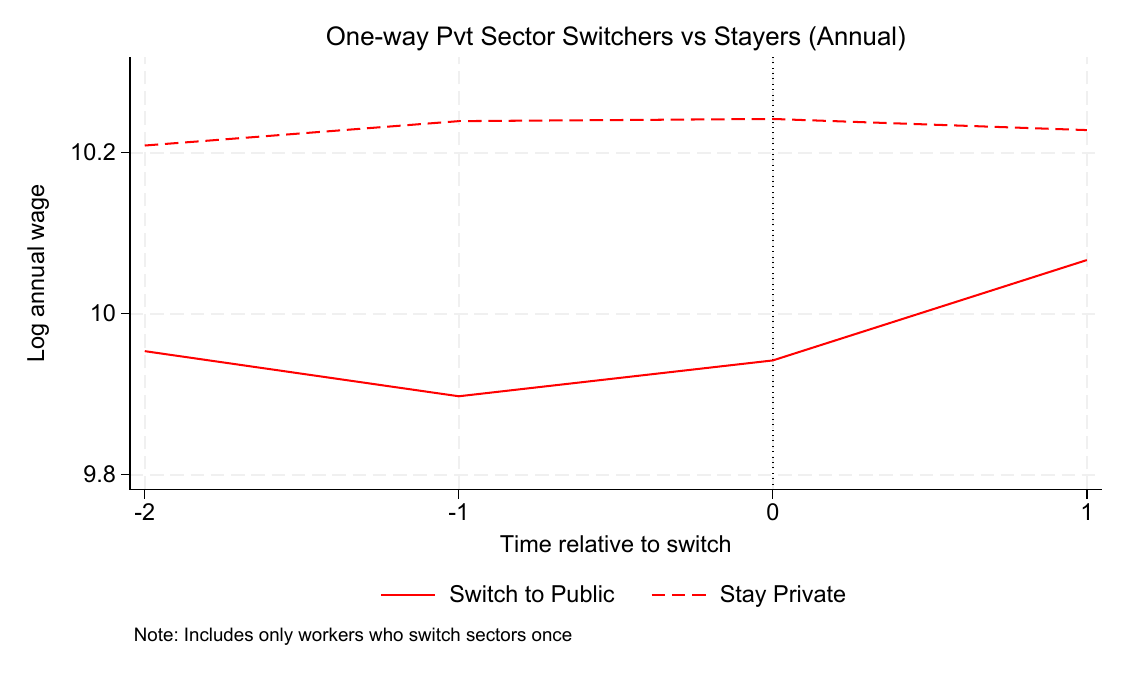}
        \caption{Annual Wage Changes - Private Sector Switchers and Stayers}
        \label{fig:annual_pvt}
    \end{subfigure}
    \caption{Distributions of Hourly and Annual Wage Changes}
    \label{fig:wage_changes}
\end{figure}

         \noindent Table \ref{tab_trans_proba_obs} above highlights how rare sector-switching is, particularly when considering full-time employment. Year-on-year a mere 1.2\% switch sectors in full-time jobs. 
         In prior works on the subject leveraging fixed-effects, identification relies on the assumption that sector switching is random - more specifically, that changes in wage are orthogonal to switching sectors. This assumption appears plausible when examining hourly wage changes, as shown in figures \ref{fig:hourly_pub} and \ref{fig:hourly_pvt} 
         , where the wage distribution across sectors shows limited patterns of systematic variation among movers. The pre-trend of stayers and movers is similar. However, when analyzing annual wage changes, a different narrative emerges. Figures \ref{fig:annual_pub} and \ref{fig:annual_pvt} suggest that sector switching is systematically associated with an increase in earnings in both directions (although a smaller increase when transitioning into the public sector). Here, we can observe a strong Ashenfelter dip in earnings for switchers. This pre-trend of falling wages for switchers whose earnings then increase on switching sectors suggests that sector-switching is not orthogonal to wages. This discrepancy between hourly and annual wages gives reason to revisit the assumption of randomness, as movers may be selecting sectors based on broader income considerations than hourly wage rates alone. 

            \subsection{Women and men move differently in the labor market} \label{subsec_diffmoves}
            \noindent Table \ref{tab_trans_proba_obs} presents transition probabilities across employment states, comparing non-employment, private, and public sector (full-time and part-time) employment segmented by gender\footnote{Note that these figures imply an inactivity rate of about one-third (referred to as non-employment in this study), and private:public sector employment ratio of approximately 75:25 as share of all employment. These figures are in line with national statistics published \href{https://www.insee.fr/fr/statistiques/3676623?sommaire=3696937}{here} and \href{https://www.insee.fr/en/statistiques/4997371}{here}.}. Key findings are: first, the higher persistence in employment status within both public and private sectors, as evidenced by high probabilities for remaining in the same employment type (e.g., 0.89 for private full-time and 0.93 for public full-time).

\begin{table}[ht!] 
    \centering
    \scriptsize
    \setlength{\tabcolsep}{4pt} 
    \renewcommand{\arraystretch}{1.1} 
    \caption{Transition Probabilities (Aggregate, Men, Women)}
    \begin{tabular}{@{}lccccc@{}}
    \toprule
    \textbf{From/To}     & NE & Pvt FT & Pub FT & Pvt PT & Pub PT \\
    \midrule
    \multicolumn{6}{c}{\textbf{Aggregate}} \\
    \midrule
    Non-employment (NE)       & \centering 0.82  & 0.11       & 0.02      & 0.04      & 0.01      \\
    Private full-time (Pvt FT) & \centering 0.08  & 0.89       & 0.002     & 0.03      & 0.0004    \\
    Public full-time (Pub FT)  & \centering 0.02  & 0.006       & 0.93      & 0.0015     & 0.04      \\
    Private part-time (Pvt PT) & \centering 0.17  & 0.12       & 0.004     & 0.7       & 0.005      \\
    Public part-time (Pub PT)  & \centering 0.10  & 0.01       & 0.13      & 0.01      & 0.75      \\
    \textit{Total} & \centering \textit{0.31 }& \textit{0.42} & \textit{0.13} & \textit{0.10} &\textit{ 0.04} \\
    \bottomrule
    \end{tabular}
    \vspace{1em} 
    \begin{tabular}{@{}lccccc|ccccc@{}}
    \toprule
    & \multicolumn{5}{c}{\textbf{Men}} & \multicolumn{5}{c}{\textbf{Women}} \\
    \cmidrule(lr){2-6} \cmidrule(lr){7-11}
    \textbf{From/To} & NE & Pvt FT & Pub FT & Pvt PT & Pub PT & NE & Pvt FT & Pub FT & Pvt PT & Pub PT \\
    \midrule
    NE       & \centering 0.83 & 0.13 & 0.01 & 0.02 & 0.005 & 0.81 & 0.08 & 0.02 & 0.06 & 0.02 \\
    Pvt FT   & \centering 0.08 & 0.90  & 0.002 & \textcolor{red}{\textbf{0.02}} & 0.0003 & 0.09 & 0.86 & 0.003 & \textcolor{red}{\textbf{0.05}} & 0.001 \\
    Pub FT   & \centering 0.02 & 0.006 & 0.95 & 0.001 & \textcolor{red}{\textbf{0.02}} & 0.03 & 0.006 & 0.92 & 0.002 & \textcolor{red}{\textbf{0.05}} \\
    Pvt PT   & \centering 0.22 & \textcolor{blue}{\textbf{0.22}} & 0.004 & \textbf{0.55} & 0.004 & 0.15 & \textcolor{blue}{\textbf{0.09}} & 0.004 & \textbf{0.74} & 0.01 \\
    Pub PT   & \centering 0.15 & 0.02 & 0.15 & 0.01 & \textbf{0.67} & 0.09 & 0.01 & 0.13 & 0.01 & \textbf{0.76} \\
    \textit{Total} & \centering \textit{0.31 }& \textit{0.53} & \textit{0.10} & \textit{0.05} &\textit{ 0.01}& \textit{0.31} & \textit{0.31} & \textit{0.16} &\textit{ 0.16} & \textit{ 0.06}\\
    \bottomrule
    \end{tabular}
    \label{tab_trans_proba_obs}
\end{table}

 \noindent Notably, men exhibit greater stability in private full-time roles (0.90) compared to women (0.86). Second, \textcolor{red}{transition rates into part-time roles show a gender disparity}: women are more than twice as likely as men to move into part-time jobs (0.05 vs. 0.02) from full-time positions in both sectors. They also \textbf{remain in part-time roles more often}, with \textcolor{blue}{less than half the probability of transitioning to private full-time jobs compared to men}. Men are more likely to move from non-employment to private full-time roles, underscoring gender disparities in workforce re-entry.

            \subsubsection{The Part-Time Paradox: Women's Dominance in Flexible Work} \label{subsec_parttime}
            \noindent A significant advantage of using the DADS data is that it allows us to observe the hours workers are compensated for, including paid time off. In other data sources, respondents are asked how many hours they work on average, or in that week, like in the Labor Force survey. In a study like this, where we want to be able to predict transitions and wages over a lifetime, having accurate and administrative records is crucial to avoid recall bias and measurement errors that can arise from self-reported data. This precision ensures a more reliable estimation of lifetime earnings trajectories and sectoral transitions, which are key to understanding long-term labor market dynamics.             
            
           \noindent Women's labor supply significantly differs from men's across public and private employment sectors, as evidenced by disparities in compensated hours captured through administrative data. Table \ref{tab:gender_hours_diff} highlights key differences in average annual compensated hours between men and women across employment types. In both full-time public and private sectors, men consistently have more compensated hours on average than women, with the gap being more pronounced in the private sector. For example, in the full-time private sector, men are compensated for 63.57 more hours annually than women (roughly translates to two weeks), while in the public sector, this difference narrows to 20.65 hours. These trends suggest that the public sector offers more equitable compensation structures for women compared to the private sector.
 \begin{table}[ht]
    \centering
    \footnotesize
    \caption{Gender Differences in Compensated Hours by Sector and Employment Type}
    \begin{tabular}{llcccccc}
        \toprule
        & & \multicolumn{2}{c}{Observations} & \multicolumn{2}{c}{Average Annual Hours} & \multicolumn{1}{c}{Difference} \\
        \cmidrule(r){3-4} \cmidrule(r){5-6}
        Type & Sector & Men & Women & Men & Women & \\
        \midrule
        Full-time & Private & 552,026 & 315,664 & 1,833.399 & 1,769.83 & 63.569*** \\
                  & Public  & 107,911 & 165,376 & 1,817.262 & 1,796.615 & 20.647*** \\
        Part-time & Private & 51,769  & 170,255 & 1,111.697 & 1,226.44  & -114.743*** \\
                  & Public  & 10,546  & 64,053  & 1,110.931 & 1,287.2   & -176.27*** \\
        \bottomrule
    \end{tabular}\\
    \tiny \textbf{Note:} The last column "Difference" is the difference in compensated hours between and women for each employment state and the *s denote the significance of t-tests (with uneqal variances) for these differences.
    \label{tab:gender_hours_diff}
\end{table}

             \noindent Women's labor supply patterns reveal a striking paradox across public and private employment sectors. Table \ref{tab:gender_hours_diff} illustrates that while men consistently have more compensated hours on average in full-time roles, this trend reverses dramatically for part-time work. In the full-time private sector, men are compensated for 63.57 more hours annually than women. However, in part-time roles, women significantly outpace men in compensated hours, with the most pronounced difference in the public sector where women log 176.27 more hours annually than their male counterparts. This part-time paradox is further emphasized by the prevalence and persistence of part-time work among women: it accounts for 32\% of women's employment compared to just 9\% for men, and women are less likely to transition out of these roles. These findings suggest that part-time work, particularly in the public sector, may be serving as a long-term strategy for women to balance work and other responsibilities, rather than a temporary state as it appears to be for men. Women's compensation within part-time employment is higher both in terms of annual and hourly wages in the public sector. A breakdown by education levels further reveals that the patterns in compensated hours persist across all categories (with higher educational attainment mitigating the gender gap in hours), with the public sector showing smaller differences in full-time compensated hours between men and women and larger differences in part-time compensated hours.
            


\subsection{Differences in earnings are larger in the private sector, especially for educated workers} \label{subsec_paygaps}
The analysis of wage differentials in France reveals persistent pay gaps between men and women that are more pronounced in the private sector, particularly for highly educated workers. Using OLS regressions on log wages, women earn 14.3\% less in annual wages and 14.7\% less in hourly wages compared to men, controlling for other factors (Table \ref{tab:combined_regs}). However, this gap narrows in the public sector, as evidenced by the positive interaction between female and full-time public sector employment (5.4\% for annual wages and 3.8\% for hourly wages) extracted from an alternate specification\footnote{Refer to table \ref{tab:regression_results_extra} in the appendix.}. Figure \ref{fig:combined} which graphs mean wages by sector and sex alludes to these differences in annual pay across the sectors by gender. The trajectory of men in the private sector is, for the most part, indistinguishable from those in the public sector as there is a significant overlap. Conversely, for women, there is no overlap in the public and private sector earnings as the public sector trajectory lies squarely above that of the private sector.

\begin{figure}[H]
    \centering
    \begin{subfigure}[b]{0.45\textwidth}
        \centering
        \includegraphics[width=\textwidth]{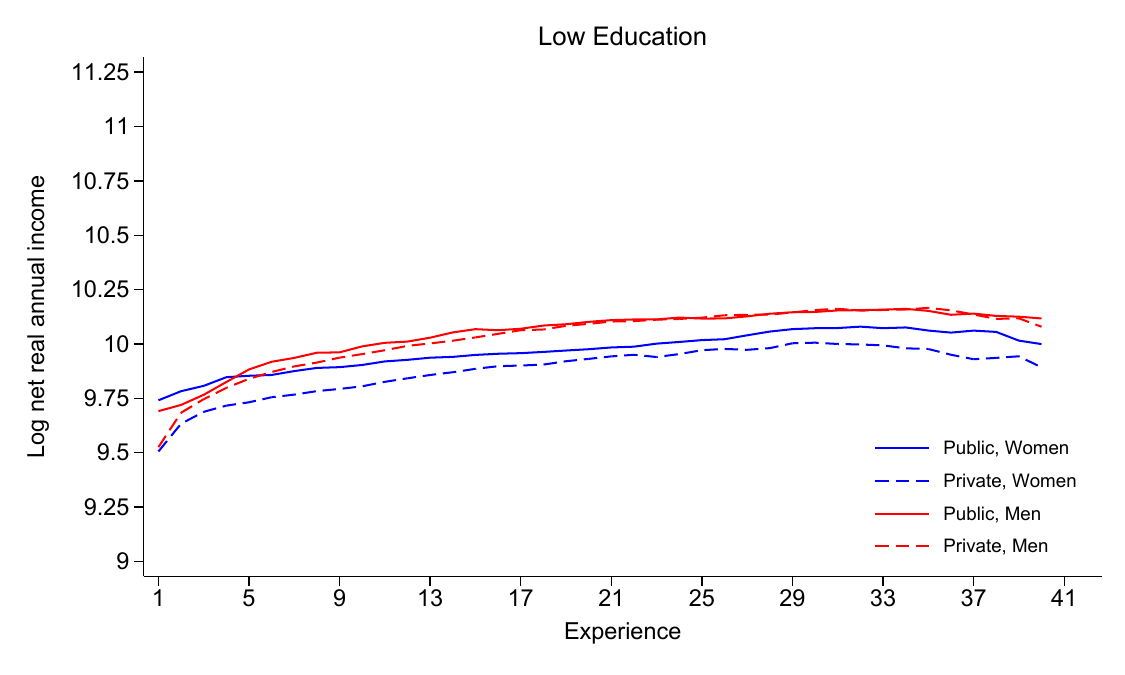}
        \caption{Low Education}
    \end{subfigure}
    \hfill
    \begin{subfigure}[b]{0.45\textwidth}
        \centering
        \includegraphics[width=\textwidth]{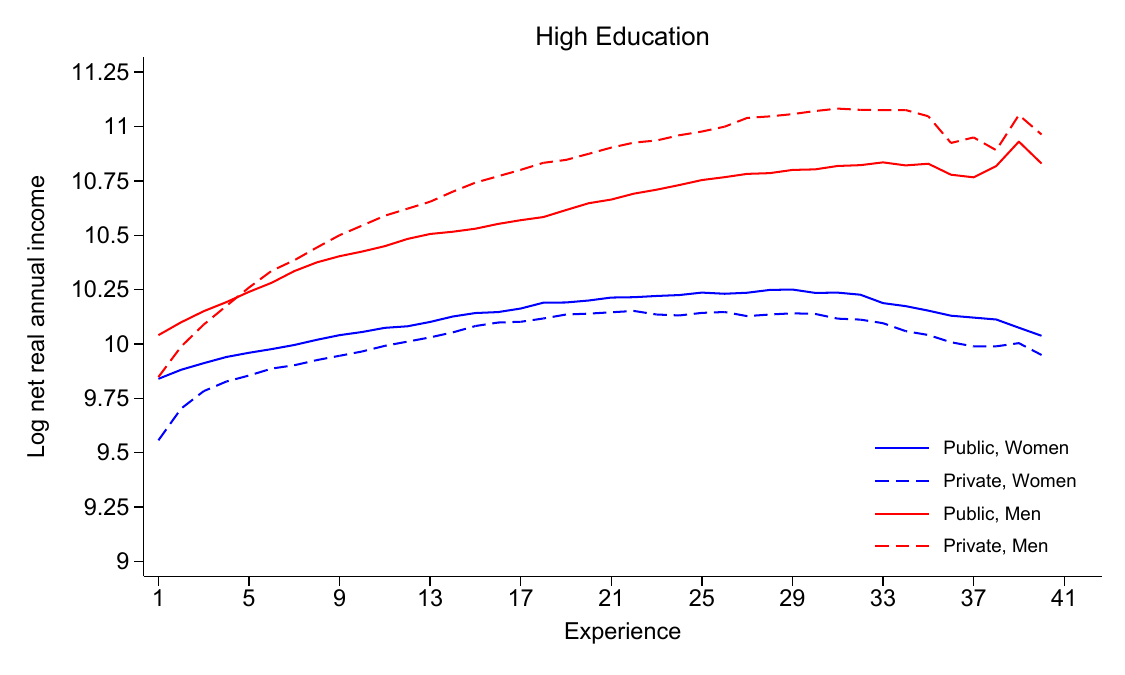}
        \caption{High Education}
    \end{subfigure} 
      \caption{Annual wage dispersion by sector, sex and sector,  over experience}
      \label{fig:combined}
\end{figure}

\begin{table}[htbp]
\centering
\footnotesize
\caption{Key Regression Results (Log Wages)}
\label{tab:combined_regs}
\scalebox{0.8}{
\begin{tabular}{lcc|cc|cccc}
& \multicolumn{2}{c|}{OLS} & \multicolumn{2}{c|}{Panel Fixed effects} &\multicolumn{4}{c}{Panel Fixed Effects by Gender} \\
\cline{2-3} \cline{4-5} \cline{6-9}
Variable & Annual & Hourly & Annual & Hourly & Men & Men & Women & Women \\
& Wage & Wage & Wage & Wage & Annual & Hourly & Annual & Hourly \\
\hline
Female & -0.143*** & -0.147*** & & & & &&\\
 & (0.001) & (0.001) & & & & \\
Full-time Public & -0.025*** & -0.047*** &0.087*** &0.016*** & -0.013*** & -0.013*** & 0.148*** & 0.023*** \\
 & (0.001) & (0.001) & (0.003)& (0.001)& (0.004) & (0.002) & (0.004) & (0.002) \\
Part-time Private & -0.717*** & -0.157*** & -0.370***& 0.059*** & -0.370*** & 0.097*** & -0.365*** & 0.035*** \\
 & (0.002) & (0.001) & (0.001)&(0.001) &(0.002) & (0.001) & (0.002) & (0.001) \\
Part-time Public & -0.546*** & -0.104*** & -0.221***&0.062*** & -0.470*** & 0.046*** & -0.129*** & 0.065*** \\
 & (0.002) & (0.001) & (0.003)& (0.001)& (0.005) & (0.003) & (0.004) & (0.002) \\
Medium Education & 0.300*** & 0.269*** & & & & & & \\
 & (0.001) & (0.001) & & & & \\
High Education & 0.545*** & 0.531*** & & & & & & \\
 & (0.001) & (0.001) & & & & \\
Experience (per decade) & 0.459*** & 0.308*** & -0.305***& 0.536*** & 0.637*** & 0.460*** & 0.546*** & 0.355*** \\
 & (0.002) & (0.001) & (0.009) & (0.005)& (0.003) & (0.002) & (0.004) & (0.002) \\
Experience Squared & -0.082*** & -0.049*** &-0.081*** &-0.053*** & -0.107*** & -0.062*** & -0.074*** & -0.041*** \\
 & (0.000) & (0.000)& (0.001)&(0.000) & (0.001) & (0.000) & (0.001) & (0.000) \\
\hline
Observations & 1,441,964 & 1,437,538 & 1,441,964 & 1,437,538 &725,048 & 722,229 & 716,916 & 715,309 \\
Number of individuals & 248,084 & 247,586  & 248,084 & 247,586 & 123,663 & 123,321 & 124,421 & 124,265 \\
Year fixed effects & \checkmark & \checkmark & \checkmark & \checkmark & \checkmark & \checkmark & \checkmark & \checkmark \\
Rho & & & 0.938&0.925 & 0.870 & 0.901 & 0.828 & 0.898 \\
\hline
\multicolumn{7}{l}{Standard errors in parentheses} \\
\multicolumn{7}{l}{*** p<0.01, ** p<0.05, * p<0.1} \\
\end{tabular}
}
\end{table}

\noindent Part-time employment significantly impacts wages, with substantial penalties observed. Part-time private sector workers earn 71.7\% less in annual wages and 15.7\% less in hourly wages relative to full-time private sector workers. However, public part-time employment appears to offer better wage outcomes than private part-time employment. The panel fixed effects regression results presented in Table \ref{tab:combined_regs} show that while both types of part-time work are associated with lower annual wages compared to full-time private sector work, the penalty is smaller for public part-time work (-12.9\% for women) compared to private part-time work (-36.5\% for women). The high rho indicates that most of the variance is due to individual fixed effects rather than idiosyncratic errors. Moreover, public part-time work is associated with modestly higher hourly wages (6.5\% increase for women) compared to full-time private sector work, while private part-time work shows a smaller hourly wage premium (3.5\% for women) - the slightly higher wage rates for part-time work is possibly a compensating differential.

\noindent Education plays a crucial role in wage determination, with higher education associated with substantial wage premia. Medium education increases wages by about 30\%, while high education increases wages by approximately 54\% compared to low education. However, the returns to education differ between sectors and genders. 
Experience contributes significantly to wage differentials, with wages increasing at a decreasing rate. The coefficient for experience (per decade) is 45.9\% for annual wages and 30.8\% for hourly wages, with negative squared terms indicating diminishing returns. The public sector shows slightly higher returns to experience, with an additional 1.4\% increase per decade for annual wages. While the public sector appears to offer a more equitable environment, especially for women and those with more experience, substantial gender pay gaps persist across both sectors, particularly for highly educated workers in the private sector. The advantage of public sector employment is especially pronounced for part-time workers, offering better wage outcomes compared to private sector part-time employment.

	\section{Model}\label{model_section}
    The methodology in this study adapts the framework developed in a seminal paper on the public pay gap in Britain \citet{ppgap}, later extended in \citet{ppgapeur}. The components directly borrowed from those papers are detailed in Appendix \ref{app_model}. For clarity and readability, the main likelihood equations (\ref{lhood} and \ref{llhoodmax}) and the lifetime earnings equation (\ref{lifetime}) are retained in the main text, even though they closely mirror those in the earlier work. Likewise, the model structure outlined in Section \ref{section::5.1} follows a similar naming convention but differs in several key components.

     The main methodological adaptation in this paper lies in the specification of the income process. Whereas the earlier studies used an AR2 framework with restrictive sample selection, log earnings here are modeled as a more flexible AR1 process. This approach avoids sample selection restrictions, captures the persistence of log wages, and allows for the influence of both stochastic noise and covariate-driven variation.

       The resulting dynamic model jointly estimates income trajectories, employment dynamics, and sectoral choice, while incorporating unobserved heterogeneity. The details on the sequential EM used is detailed in section \ref{section::5.5}. Individuals are allowed to differ not only in their earnings levels and earnings mobility, but also in their propensity to enter non-employment or to work in the public sector, conditional on fixed characteristics such as labor market experience, education, and gender. These latent patterns give rise to distinct income and mobility classes within the data. Lifetime values of jobs are then constructed for each sector and analyzed comparatively across states, transition classes, earnings groups, and gender.


	\subsection{Structure}\label{section::5.1}
	The data is structured as follows: For each individual $i \in (1,..., N)$, we have a record of respective income flows, employment states, fixed characteristics, denoted by a vector $\boldsymbol{x}_i=(\boldsymbol{y}_i, \boldsymbol{S}_i, \boldsymbol{z}_i^v, \boldsymbol{z}_i^f)$ over a time period $T_i$. Each boldface variable here represents a vector.
	    \begin{itemize}
	        \item $\boldsymbol{y}_i=(y_{i1},...,y_{iT_i})$ is the sequence of individual $i$'s log income (real net annual wage) at each time period $t$ where $t \in (1,2,,...,T_i)$ 
	        \item $\boldsymbol{S_i} =(S_{i1},..., S_{iT_i})$ short for state, records the sectoral state of the worker. The different states are defined as follows for a person $i$ in the $t^{th}$ year:
	        \begin{itemize}
    	            \item ${S}_{it}=0$ : not employed.
    	            \item ${S}_{it}=1$ : employed in a full-time private sector job.
    	            \item ${S}_{it}=2$ : employed in a full-time public sector job.
    	            \item ${S}_{it}=3$ : employed in a part-time private sector job.
    	            \item ${S}_{it}=4$ : employed in a part-time public sector job.
	        \end{itemize}
            \item $\boldsymbol{z}^v_i=(z^v_{i1},..., z^v_{iT_i})$ is the sequence of time-varying individual characteristics. In this analysis labor market experience over a decade (defined as the cumulative duration of employment of individual $i$) and its square are considered. Note that conditional on the $state$ at time $t$, $z_i^v$ is deterministic.
    	    \item $\boldsymbol{z}^f_i$ is the set of individual fixed characteristics. It includes highest academic qualification, labor market experience when first observed in the sample, and gender.  $\boldsymbol{z}^v_i$ is thus deterministic conditional on $\boldsymbol{z}^f_i$.
    	\end{itemize}
    In addition to observed individual heterogeneity captured by $\boldsymbol{z}^v_i$ and $\boldsymbol{z}^f_i$, there might exist certain unobserved individual characteristics which may influence wages or movement or selection into the various labor market states. Thus we should supplement the data vector $\boldsymbol{x}_i$ by appending a set $\boldsymbol{k}_i$ of such (time-invariant) unobserved characteristics. 
    The aim of the model is to simultaneously estimate transitions between the aforementioned labor market states and income trajectories within and between employment sectors. To this end, individual contributions to the $complete$ likelihood—i.e. the likelihood of $(\boldsymbol{x}_i,\boldsymbol{k}_i)$, including unobserved variables is described as:
        \begin{equation}\label{lhood}
         \mathcal{L}_i(\boldsymbol{x}_i,\boldsymbol{k}_i)=\ell_i(\boldsymbol{y}_i|\boldsymbol{S}_i,\boldsymbol{z}_i^f,\boldsymbol{k}_i)\cdot\ell_i(\boldsymbol{S}_i|\boldsymbol{z}_i^f,\boldsymbol{k}_i)\cdot\ell_i(\boldsymbol{k}_i|\boldsymbol{z}_i^f)\cdot\ell_i(\boldsymbol{z}_i^f)
	    \end{equation} 
    Through each individual decomposition in equation \eqref{lhood}, the aim is to capture the distribution of observed individual characteristics; the distribution of unobserved individual heterogeneity given observed characteristics; and the likelihood of individual earnings and labor market state trajectories given individual heterogeneity. 
    Then, estimates are obtained from parameters by maximizing the sample log-likelihood
	     \begin{equation}\ \label{llhoodmax}
	        \sum_{i=1}^N \ln \Bigg( \sum_{k_i^m=1}^{K^m} \sum_{k_i^y=1}^{K^y} \mathcal{L}_i(\boldsymbol{x}_i,\boldsymbol{k}_i) \Bigg)
	    \end{equation}

\par The next sub-sections delve deeper into the individual components of the individual likelihood equation \eqref{lhood} and elaborate on their implications and functional forms.
\subsection{Unobserved heterogeneity}\label{section::5.2}
\par Starting from the third term of the likelihood function $\mathcal{L}_i(x_i,k_i)$ given by equation \eqref{lhood}, two types of unobserved heterogeneity, $k_i=(k_i^y,k_i^m)$ is considered. This is motivated by the evidence presented in subsections \ref{subsec_diffmoves} and \ref{subsec_paygaps} that men and women move differently in the labor market and thus earn different wages.
    \begin{itemize}
         \item $k_i^m$ : \textbf{Transition class}. This accounts for heterogeneity of individuals in terms of propensity to be in the different non-/employment states, $S_i$. $k_i^m$ conditions the parameters related to employment and sectoral histories.
	    \item $k_i^y$ : \textbf{Income class.} This accounts for heterogeneity in terms of income. $k_i^y$ conditions the income parameters relating to individuals' income process.
    \end{itemize}
Both these heterogeneity classes are time-invariant random effects for an individual. The transition class, \textbf{$k_i^m$} is conditional on the fixed individual characteristics and helps in modelling the selection of individuals into the various employment states, $S_i$. $k_i^y$ on the other hand, being time invariant as well, helps to increase the persistence of income ranks of an individual. In essence, we are using a finite mixture approach here to model unobserved heterogeneity of individuals, where each individual can belong to one of $K^m$ transition classes and $K^y$ income classes. The total number of possible classes is thus $K=K^m \times K^y = 12$. 
Details on estimation of unobserved heterogeneity, transition classes, and the number of classes chosen is presented in the appendix sections \ref{app_unob_het} and \ref{section::5.3}.

\subsection{Income process and Income class: \texorpdfstring{$K^y$}{Ky}}\label{section::5.4}
The first term of the individual likelihood equation \eqref{lhood} models the likelihood of the individual's income path over the years. Assume $y_{it}$, the log-income of individual $i$ in year $t$ to be the realization of a first-order Markov process. The likelihood of an individual's income trajectory can thus be expanded as:
\begin{equation}\label{lhoody}
    \ell(\textbf{y}) = \ell(y_1) \cdot \prod_{t=2}^{T}\ell(y_t|y_{t-1}) \\
    = \ell(y_1) \cdot \prod_{t=2}^{T}\frac{\ell(y_t,y_{t-1})}{\ell(y_{t-1})}
\end{equation}
Now each of these yearly log-income streams, are assumed to be normal, and conditional on current labor market state and observed and unobserved heterogeneity. 
\begin{center}
$y_{it} | S_{it},z_{it}^v,z_i^f,k_i^y \sim \mathcal{N}(\mu_{it},\sigma^2_{it})$
\end{center}

\begin{center}
    with $\mu_{it} = \mu(S_{it},z_{it}^v,z_i^f,k_i^y)$ and  $\sigma_{it} = \sigma(S_{it},z_{it}^v,z_i^f,k_i^y)$
\end{center}
We model the $\mu_{it}$ using an OLS regression of $y_{it}$ over the current labor market state and observed and unobserved heterogeneity.
 \begin{equation}\label{mu}
    \mu(S_{it},z_{it}^v,z_i^f,k_i^y) = \begin{pmatrix}
    S_{it} & z_{it}^v & z_i^f & k_i^y & S_{it}*k_i^y
    \end{pmatrix}
    ' \cdot  \mu
\end{equation}
Here the $*$ represents the interaction term of two variables. 

 \begin{equation}\label{sigma}
    \sigma(S_{it},z_{it}^v,k_i^y) = \sqrt{ \exp \Bigg[\begin{pmatrix}
    S_{it} & z_{it}^v & k_i^y & S_{it}*k_i^y
    \end{pmatrix}
    '\cdot  \sigma \Bigg] }
\end{equation}   

\noindent The functional form of $\sigma(\cdot)$ constrains the standard deviation of the log-income $y_{it}$ to be positive. The effect of the fixed individual heterogeneity $z_i^f$ is subsumed through its link to $k_i^y$. Using $y_{it}$, $\mu$ and $\sigma$, the normalized log-income is constructed as $\Tilde{y_{it}}=\frac{y_{it}-\mu_{it}}{\sigma_{it}}$. Therefore, $\Tilde{y_{it}} \sim \mathcal{N}(0,1)$.

The pair $(\Tilde{y}_{it}, \Tilde{y}_{i,t-1})$ is a vector which follows a bivariate multinomial normal distribution. It has a covariance matrix $\underline{\tau}_{it}^{(2)}$ which can be expanded as:
 \begin{equation}\label{covmat}
    \underline{\tau}_{it}^{(2)} = \begin{pmatrix}
    1 & \tau_{i,t,t-1}\\
    \tau_{i,t,t-1} & 1
    \end{pmatrix}.
\end{equation}
 $\tau$ is individual-specific and is allowed to vary with observed and unobserved  heterogeneity and with employment $state$ at $t\text{ and } t-1$. We therefore create a function, $\tau_1(\cdot)$,
 \begin{equation}\label{tau}
    \begin{split}
         \tau_{i,t,t-1} = \tau_1 \big(S_{it},S_{i,t-1},z_{it}^v,z_i^f,k_i^y,k_i^m  \big).
    \end{split}
\end{equation}
The functional form of $\tau_1(\cdot)$ is: 
     \begin{equation}\label{tau1}
     \tau_1(S_{it},S_{i,t-1},z_{it}^v,k_i^y) = -1+2 \cdot \Lambda \Bigg[ \begin{pmatrix}
    S_{it} & S_{i,t-1} & z_{it}^v & k_i^y & k_i^m & S_{it}*k_i^y & S_{i,t-1}*k_i^y
    \end{pmatrix} \cdot \zeta \Bigg]
    \end{equation}
 Here $\Lambda(x)={(1+e^{-x})}^{-1}$. The functional form of $\tau_1(\cdot)$ includes a transformation in the form of $-1+2\cdot \Lambda(\cdot)$. This constrains the correlation coefficient between normalized incomes at year t and t-1 to lie between [-1,1]. Likelihood of an individual's income trajectory $\mathbf{y_i}$ defined earlier in equation \eqref{lhoody} is:
\begin{equation*}
    \begin{split}
         \ell(\textbf{y}) = \ell(y_1) \cdot \prod_{t=2}^{T}\frac{\ell(y_t,y_{t-1})}{\ell(y_{t-1})} \\
    \end{split}
\end{equation*}

 \noindent The numerator in this expression can be written as a joint density of a pair of normalized log-earnings, $ \Tilde{y}_{it}=\frac{y_{it}-\mu_{it}}{\sigma_{it}}$

 \begin{equation}
     \ell(y_t,y_{t-1})=\dfrac{1}{\sigma_t \sigma_{t-1}} \cdot \varphi_2 \Big( \Tilde{y}_{it}, \Tilde{y}_{i,t-1}; \underline{\tau}_{it}^{(2)}\Big) 
 \end{equation}

\noindent where $\varphi_2 ( \cdot ; \underline{\tau}^{(2)})$ is the bivariate normal pdf with mean 0 and covariance matrix $\underline{\tau}^{(2)}$\footnote{Note that $\varphi_2(\cdot)$ is defined as: 
\begin{equation} \label{pdf}
    \varphi_n(\boldsymbol{Y};\Sigma)=\ddfrac{1}{\sqrt{(2\pi)^2\Delta_{\Sigma}}}\cdot exp({-\frac{1}{2}\cdot(\boldsymbol{Y-\mu})'\Sigma^{-1}(\boldsymbol{Y-\mu})})
\end{equation}}. Substituting this expression to equation \eqref{lhoody} gives us:
\begin{equation}
    \begin{split}
    \ell(\textbf{y}) = \ell(y_1) \cdot \prod_{t=2}^{T}\frac{\ell(y_t,y_{t-1})}{\ell(y_{t-1})} \\
     = \dfrac{\varphi_1(\Tilde{y}_1)}{\sigma_1} \cdot \Bigg( \prod_{t=2}^T \dfrac{\varphi_2(\Tilde{y}_{t},\Tilde{y}_{t-1}; \tau_t^{(2)})}{\varphi_1(\Tilde{y}_{t-1})} \cdot \prod_{t=2}^T \dfrac{1}{\sigma_{t} } \Bigg)\\
    = \Bigg(\prod_{t=1}^T \dfrac{1}{\sigma_{t} }\Bigg) \cdot  \ddfrac{ \prod_{t=2}^T \varphi_2(\Tilde{y}_{t},\Tilde{y}_{t-1}; \tau_t^{(2)})}{\prod_{t=2}^{T-1} \varphi_1(\Tilde{y}_{t})}  \\
    \end{split}
\end{equation}

\noindent With these specifications in place, normalized incomes are thus assumed to follow an AR(1) process:
 \begin{equation}\label{yprocess}
    \Tilde{y}_{it}= \rho_{it} \cdot \Tilde{y}_{i,t-1} + \epsilon_{it}
\end{equation}
where the innovations $\epsilon_{it}$ are normal with mean 0 and serially uncorrelated.

\subsection{Linking \texorpdfstring{$\rho$}{rho} to \texorpdfstring{$\tau$}{tau} and \texorpdfstring{$\sigma^2$}{sigma\^2}}

Equation \ref{yprocess} describes the AR(1) process for the normalized wage. We now resort to the standard relations between these parameters in a typical AR(1) process. We recall, in a first order autoregressive process, the autocovariance at lag 1,$\tau_1$ is related to the $\sigma$ and the autocorrelation coefficient $\rho$ as: 
\begin{equation}\label{tausigrho}
    \tau_1=\frac{\sigma^2}{1-\rho^2}
\end{equation}
Simplifying and solving for $\rho$, we get
\begin{equation}\label{rhosigtau}
    \rho = \frac{-\sigma^2 + \sqrt{\sigma^4 + 4\tau^2}}{2\tau},
\end{equation}
Now that we have $\rho$, we can recreate the income process based on the first wage and Equation \ref{yprocess}. By substituting the equation for normalized log-wages into the AR(1) process, we can return to the original (unnormalized) process. Specifically, using $\tilde{y}_{it} = \frac{y_{it} - \mu_{it}}{\sigma_{it}}$:
\begin{equation}
y_{it} = \mu_{it} + \sigma_{it} \cdot \left( \rho_{it} \cdot \frac{y_{i,t-1} - \mu_{i,t-1}}{\sigma_{i,t-1}} + \epsilon_{it} \right).
\end{equation}
\noindent This equation describes how log-wages at time $t$ depend on:
\begin{enumerate}
    \item The individual-specific mean $\mu_{it}$, which reflects systematic factors such as experience, education, and sector,
    \item The deviation from the mean at the previous period, $(y_{i,t-1} - \mu_{i,t-1})$, weighted by the persistence parameter $\rho$,
    \item A stochastic noise term, $\sigma_{it} \cdot \epsilon_{it}$, capturing unobserved factors.
\end{enumerate}

This approach ensures that the model incorporates the underlying persistence of log-wages while accounting for the role of noise and covariate-driven variation.

\section{Model Fit}\label{estimation_section}
    \subsection{Understanding the latent classes}
	The EM algorithm was run until convergence (distance of parameters $<10^{-3}$) in order to arrive at the final estimates of the $\kappa^m$, $\chi$, $\chi_0$, $\mu$, $\sigma$, $\xi$, $\kappa^y$ coefficients. These final estimates of the coefficients are reported in the appendix table \eqref{tab:km-table} through \eqref{tab:tau-table}. 
    
    \noindent The estimation procedure identifies distinct latent transition and income classes, capturing the heterogeneity in employment dynamics and wage trajectories across the public and private sectors. The classification reveals four transition classes (related to employment sector mobility) and three income classes (characterizing wage persistence and mobility), jointly shaping career outcomes. We first look at the mobility estimations for our panel data. The individuals in our panel are sorted into 4 transition classes. 42\% individuals are sorted into the Transition class 0, around 4\% in class 1, 47\% in class 2 and 7\% in class 3. Note that the order is meaningless. The transition classes have the distribution as in Table \eqref{tab:kmdist_desc}. Note that women are overrepresented in transition classes with higher public sector participation, as expected.

         \begin{table}[ht]
        \centering
        \scriptsize
        \caption{Share of individuals by Sex, Education, Transition Class and Income Class}
        \begin{tabular}{cccccc}
        \toprule
        \multirow{2}{*}{Transition Class} & Share of & \multirow{2}{*}{Female share} & \multicolumn{3}{c}{Education} \\
        \cmidrule{4-6}
        & individuals & & Low share & Med share & High share \\
        \midrule
        0 & 42\% & 65\% & 51\% & 16\% & 32\% \\
        1 & 4\% & 66\% & 71\% & 29\% & 0\% \\
        2 & 47\% & 33\% & 86\% & 14\% & 0\% \\
        3 & 7\% & 56\% & 68\% & 10\% & 22\% \\
        \toprule
        \multirow{2}{*}{Income Class} & Share of & \multirow{2}{*}{Female share} & \multicolumn{3}{c}{Education} \\
        \cmidrule{4-6}
        & individuals & & Low share & Med share & High share \\
        \midrule
        0 & 8\% & 40\% & 60\% & 13\% & 26\% \\
        1 & 32\% & 57\% & 72\% & 17\% & 11\% \\
        2 & 60\% & 47\% & 69\% & 15\% & 16\% \\
        \bottomrule
        \label{tab:kmdist_desc}
        \end{tabular}
        \end{table}
\noindent Although the estimated latent transition classes differ markedly in their composition by sex, education, and age, the transition dynamics implied by the model are remarkably similar once these observed characteristics are conditioned on. In other words, the classes capture more the differences in \textit{who} participates in different segments of the labor market rather than differences in \textit{how} individuals move between employment states over time.

The patterns of sectoral persistence and transition—such as the high stability of full-time employment in both the public and private sectors, and the relatively frequent transitions from part-time to non-employment--are close across transition classes. However, there are differing propensities in the latent classes: specifically, those in transition class 0, $k^m=0$, have a slightly higher propensity to go into public full-time roles, those in $k^m=1$ to public part-time roles, those in $k^m=2$ to private full-time roles and finally those in $k^m=3$ to private part-time roles. This is evidenced by the multinomial logit parameters in table \ref{tab:km-table}.

\noindent Unobserved heterogeneity plays a fundamental role in explaining wage dynamics in our dataset. The latent income classification ($k^y$) uncovers distinct subpopulations characterized by different wage persistence and mobility regimes that go beyond observed demographics and labor market states. These income classes capture hidden variation in earnings trajectories that cannot be fully attributed to measured covariates like education, age, or sectoral affiliation.

The model’s goodness-of-fit notably improves with the introduction of these latent income classes, reflecting important wage heterogeneity within transition classes ($k^m$). For instance, the highest income class ($k^y=0$), comprising about 8\% of individuals, is marked by stable high earnings and high education levels, while the largest income class ($k^y=2$, 60\%) exhibits greater wage volatility and lower overall earnings.
\begin{figure}[H]
    \centering
    \includegraphics[width=.7\linewidth]{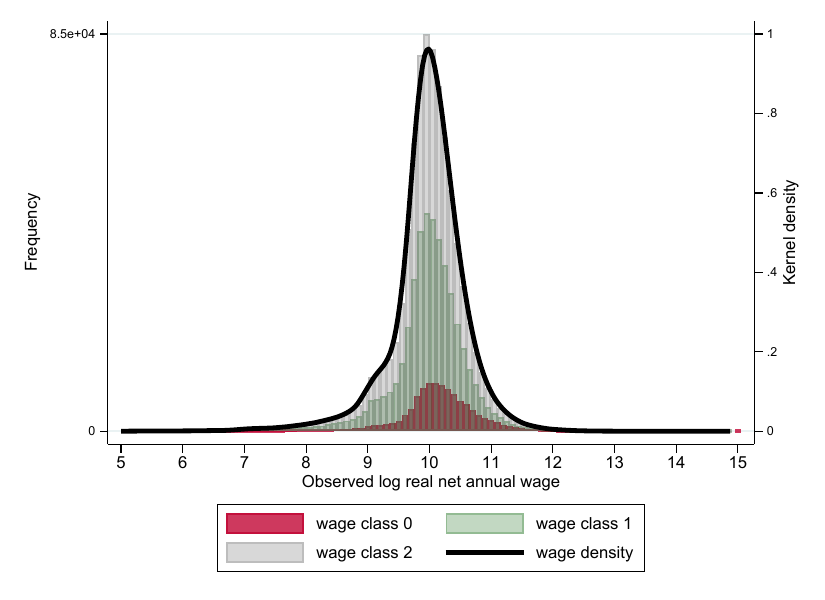}
    \caption{Observed latent wage class distribution}
    \label{fig:obs_latent_wage}
\end{figure}

Figure \ref{fig:obs_latent_wage} visually presents these distinct income distributions and highlights the stratification in wage stability and volatility across classes.

Taking a combined perspective on both income ($k^y$) and transition ($k^m$) latent classes, we observe layered unobserved heterogeneity: while the transition classes primarily capture compositional differences in employment states and demographics, the income classes reveal important heterogeneity in wage dynamics conditional on these employment patterns. This implies that unobserved heterogeneity in the model is multi-dimensional, affecting both employment transitions and wage mobility but through somewhat distinct latent structures.

This layered latent typology enriches our understanding of stratification and persistence mechanisms shaping the public-private wage gap and labor market careers. The detailed characterization of all twelve combined latent classes ($k^y \times k^m$) is compiled in the appendix (Table \ref{tab:latent_class_summary}), providing comprehensive insight into the interaction between employment stability and wage mobility across demographic groups.

While this paper employs a similar latent class framework to \cite{ppgapeur}, the findings reveal less pronounced unobserved heterogeneity in mobility classes. This difference can be attributed to the broader and more heterogeneous sample used here, which includes both men and women across five employment states (adding part-time work in both sectors). The inclusion of gender as an observed covariate absorbs substantial variation in sectoral attachment and employment stability that, in a male-only sample, would appear as unobserved mobility heterogeneity. Moreover, the richer state space—incorporating part-time employment—introduces more fluid and overlapping transition patterns, making it difficult for latent mobility classes to capture sharp behavioral distinctions. In contrast, the income classes reveal meaningful unobserved heterogeneity in wage dynamics, consistent with prior findings, suggesting that while observed factors largely explain sectoral mobility, wage trajectories retain important latent structure even after controlling for demographics and employment states.

	\subsection{Prediction}\label{section::6.3}
     As is customary, the statistical model's goodness-of-fit and predictive power were first tested in-sample presented in subsection \ref{in_sample_pred}. What has not been done in this particular literature thus far is the out-of-sample test with a data size three times as large as the training sample. This is presented in subsection \ref{out_sample_pred}.
   \subsubsection{In-sample} \label{in_sample_pred}
    Let individuals be assigned initial states for the first year that they appear in the panel, based on probabilities generated by equation \eqref{chicoeff} conditioned on individual fixed characteristics. The method of probabilistic assignment is used so as to respect the probability distributions and not directly assign the state with the highest probability \footnote{Probabilistic assignment is chosen over deterministic assignment to prioritize accuracy. Results are exactly replicable as the \texttt{set seed} command was used. Moreover, the correlation between values generated by probabilistic assignment and deterministic assignment is, on average 0.7.}. The probabilities from equation \ref{chcoeff} are used to assign the subsequent states, again employing probabilistic assignment. We can thus, test the goodness of fit and the predictive power of the model by checking year-on-year state transitions in the observed data and comparing it with the same transitions within the predicted data. We can observe quite similar patterns within these transitions as tabulated in Table \ref{tab_trans_proba_pred} compared with the earlier descriptive table \ref{tab_trans_proba_obs}. The predictive strength is corroborated by the fact that the maximum distance between the observed and predicted transition probability matrices is 0.01 on aggregate.
    
\begin{figure}[H]
        \centering
        \includegraphics[width=.7\linewidth]{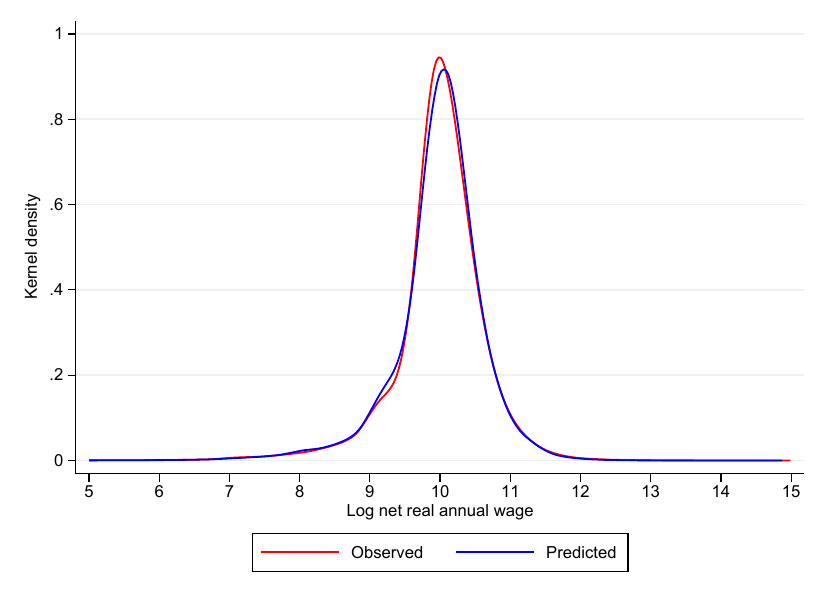}
        \caption{Observed and predicted wage densities}
        \label{fig:obs_pred_wage}
    \end{figure}

\noindent   Using the income process parameters, the log (real net annual) wages can be predicted. Once the simulation is done, we compare the simulated incomes to the observed incomes by comparing the observed and predicted wage densities as in figure \ref{fig:obs_pred_wage}. The predicted latent class composition is also tested: table \ref{tab:kmdist_desc} is almost exactly replicated. Moreover, the income/wage class composition predicted as in figure \ref{fig:pred_wage_latent} is very close to the observed figure \ref{fig:obs_latent_wage}. A word of caution here, the model slightly overpredicts full-time wages and slightly underpredicts part-time wages as shown in figure \ref{fig:density_by_state_obs_pred}. However, figure \ref{fig:obs_pred_wage} (below) and figure \ref{fig:pred_wage_latent} (in the appendix) provide strong evidence for the predictive power of this model. The actual observed and predicted income distributions are a near exact match, including the small kink towards the left of the observed income distribution.

     \subsubsection{Out-of-sample} \label{out_sample_pred}
     To test the model’s robustness beyond the estimation sample, I applied the trained coefficients to a larger out-of-sample panel comprising 829,232 individuals and 6,194,037 observations (thrice the size of the in-sample panel used). In this prediction process, I retained only each individual's first observed employment spell and used the estimated coefficients to simulate their subsequent (non-)employment trajectories and associated wages. This approach allows for assessing how well the model generalizes to a broader population while accounting for both observed and unobserved heterogeneity in labor market outcomes.

    \noindent All the state transition and wage prediction results presented in the in-sample section \ref{in_sample_pred} were also replicated for this larger panel and are included in the appendix \ref{sec_oos}. The consistency of the predicted trends across samples reinforces the model’s validity in capturing employment mobility, wages, and sectoral differences over time.

	\section{Results: Public and Gender Premia in Lifetime Earnings }\label{lifetime}
    \subsection*{Constructing Lifetime Earnings}\label{section::5.6}
Once the maximization problem is resolved and we have stable parameters of the income and job mobility parameters, simulations of employment and income trajectories for the individuals in the sample are carried out until retirement age. Assuming that, upon retiring (which is assumed to happen at age 60 denoted as $T_R$), a given individual enjoys a residual value of $V_R$, then the lifetime value as of experience level $t$ of an individual’s simulated future income trajectory $y_{s\geq t}$ is written as: 
	     \begin{equation}\label{value}
	        V_t(y_{s\geq t})=\sum^{T_R}_{s=t}\beta^{s-t}\cdot \exp(y_s)+\beta^{T_R-t}\cdot V_R
	    \end{equation}
	    where $\beta \in (0, 1)$ is the discount factor and $\exp(y_t)$ is the projected real net annual wage that the individual earns from (log) income $y_t$. We could also interpret $exp(y_t)$ as the risk-neutral utility from earnings.
The log of the above $V_t$ - the log of the discounted sum of lifetime earnings (hereon referred as "lifetime earnings") under different pathways  are calculated and compared sectorally. The discount factor $\beta = 0.95$ per annum, as is standard in the literature. The value of retirement is defined as $V_R=\frac{1-\beta^{22}}{1-\beta}\times RR \times e^{y_{T_{R}-1}} $ in both the scenarios. Assuming that after retirement, individuals receive a constant flow of income equal to RR times their last income in activity and discount this flow over a period of 22 years (time between average retirement age and average life expectancy which is 82 years, in France). The particular values of $\beta = 0.95$ and $RR = 0.4$ that were taken obviously condition the results described. A lower $\beta$ would clearly make lifetime values close to current income flows and would magnify the observed public-private pay gap (the private sector earnings being far more dispersed). An increase in $RR$ on the other hand, would only magnify the impact of the income gap at the time of retirement. For the main results presented in section \ref{lifetime}, the $RR$ used is $0.40$ using the precedent set by \cite{ppgapeur}. For robustness, all the specifications to calculate the discounted sum of lifetime earnings were also estimated with an $RR=0.7$ and separate rates for the public and private sectors: $RR_{pub}=0.75$ and $RR_{pvt}=0.71$. The results are largely consistent.

An implicit assumption in computing lifetime values, is that the prevailing economic environment is stationary and individuals anticipate aging and experiencing changes in their wage and job mobility given their current state and income level, but not the inherent model paramaters to be changing over their working life. Given the stability of the share of public sector employment and of the non-employment rate in France over the sample period, the sample time-period can be considered as an average state of the business cycle. The public premium is defined as the difference in log-lifetime values (referred to as "lifetime earnings" hereon) between the public and private sector. The 'whole sample, with selection' graph relates to predicted 'raw' differences, i.e. it plots the difference between quantiles of lifetime earnings among individuals effectively observed to hold public jobs in the initial period, and corresponding quantiles of lifetime earnings among workers observed to hold private jobs in the initial period. The whole sample graphs relate to predicted differences in lifetime earnings across sectors for all individuals in the sample, i.e. it compares income and lifetime values that each individual could potentially earn in either sector.

    In this section, I present a series of insights that stem from comparing lifetime earnings\footnote{Recall that "lifetime earnings" refers to the log of the discounted sum of future earnings as defined in equation \ref{value}.} across sectors, by gender and education, with and without selection to offer a more holistic understanding of what we perceive as the public premium, first focusing on if there is one at all. Two broad sets of counterfactuals are considered: one, is to compare the "job-for-life" lifetime earnings wherein individuals (on aggregate, and by gender, by education, etc.) are assumed to hold either a public or private sector job in every period from the time they are first observed until age 60. Two, is to compare lifetime earnings "with mobility", meaning, people can transition into any of the 5 states (non-employment, full-time employment in private sector, in public sector, part-time employment in private and public sectors) until retirement based on the transition probabilities as per their observed (gender, education, experience) and unobserved factors (latent types). The first set of counterfactuals sheds light on what the earnings gap amount to over a lifetime \textit{ceteris paribus} assuming that people are employed in every period, full-time in either sector. The second set ("with mobility") gives a more realistic and accurate picture as it estimates lifetime earnings "with mobility", taking transition patterns into account.

\subsection{The Lifetime Public Sector Premium}
\noindent A significant public sector premium exists in France, particularly for women, low-educated workers, and individuals over 45. This premium persists across most of the earnings distribution even after accounting for self-selection into sectors. An oft-encountered question here is if the public premium is just a result of selection. The short answer is no, as many studies confirm. But the selection effect indeed estimates higher and larger premia or penalties as the case may be as deconstructed herein. Figure \ref{fig:w_wo_selection_premialabel} compares the counterfactual lifetime earnings from the "job-for-life" case with the counterfactual lifetime earnings wherein individuals, conditional on selection into either sector, transition freely across employment states. Of course, these transitions are informed by the patterns that are already captured from the model conditioned on observables and unobservables. Unsurprisingly, lifetime earnings are higher if individuals hold a full-time job in either sector in every period, but we know that is rare for everyone to experience. The losses in lifetime earnings is higher in private sector due to higher transitions into non-employment and worse part-time pay compared to the public sector.

\noindent \textbf{Job-for-life case: \textit{potential} lifetime earnings differences if everyone could hold a full-time job in every period. } 
With selection there is a positive public premium until the 46th percentile on aggregate. But even without selection, there is one until the 30th percentile. To understand how much of the premium we see is just a matter of selection, we must first quantify if there is a public premium "with selection" and then, if there is one "without" and for whom, if at all. 
\begin{itemize}
    \item \textbf{"with selection"}: What happens if those observed to be in public sector full-time jobs were forced to switch to the private sector instead? \\
    \textbf{Main result:} Is a public premium observed? Yes. For whom? Women (until the 69th percentile), Low educated workers (until the 60th percentile) and older workers (over 45)  (until the 76th percentile).
    \item \textbf{"without selection"}: What happens if everyone in the sample is employed full-time in the public sector vs. the private sector? \\
    \textbf{Main result:} Is a public premium observed? Yes. For whom? Women (until the 38th percentile) Low educated workers (until the 38th percentile) and older workers (over 45)  (until the 49th percentile).
\end{itemize}

\noindent Figure \ref{fig:w_wo_selection_premialabel} illustrates the public premium in lifetime earnings for job-for-life counterfactuals, both with and without selection. The graph demonstrates that a public premium persists even when selection is eliminated, though to a lesser extent and for a smaller portion of the earnings distribution. With selection, the public premium extends up to the 46th percentile of the aggregate earnings distribution, while without selection, it reaches the 30th percentile. For women, the premium without selection is observed up to the 38th percentile, compared to the 69th percentile with selection. Similarly, for low-educated workers, the premium without selection persists up to the 38th percentile (versus 60th with selection), and for older workers over 45, it extends to the 49th percentile (compared to 76th with selection).

\noindent These findings underscore that while selection plays a significant role in amplifying the observed public premium, it is not the sole factor. The persistence of the premium even without selection indicates structural differences in compensation between the public and private sectors, particularly for certain demographic groups and at lower earnings and education levels. This analysis reinforces the conclusion that the public premium is not merely a result of selection, but reflects underlying differences in sectoral compensation structures. One might be tempted looking at figure \ref{fig:w_wo_selection_premialabel} to compare the bars across categories, for example, comparing the premia for men vs. for women to see potentially if women have a higher public premium than men, but that would be misleading. The bars in figure \ref{fig:w_wo_selection_premialabel} correspond to counterfactual earnings \textit{within} that category (the percentage difference of lifetime earnings of low educated women first observed to be in the public sector from the lifetime earnings of low educated women if forced to be in the public sector give us the "with selection" red bar). To answer the question about gender premia by sector, we must compare the lifetime earnings of men and women (section \ref{sec_gender_prem}). 

\noindent \textbf{With mobility: \textit{true} lifetime earnings differences show that losses in lifetime earnings higher in private sector.} 
    The estimated loss in earnings, calculated as the difference in (log) lifetime earnings of the job-for-life estimates and those with mobility, reveals that individuals initially observed in the private sector full-time experience greater income erosion due to higher transitions into non-employment and weaker part-time wage prospects. Converting these log differences into percentage terms, the lifetime earnings loss ranges from approximately 82\% for the lowest quantiles to 5\% for the highest quantiles in the private sector, whereas losses in the public sector are consistently lower (nearly half, throughout), ranging from 63\% to 2\%. This highlights the structural disadvantages faced by private-sector workers in earnings stability, emphasizing the role of sectoral labor mobility in shaping long-term economic outcomes. Of course, this effect is not gender-neutral either and the gaps are larger for women.
    \begin{figure}[H]
    \label{loss_lifetime}
        \centering
        \includegraphics[width=0.5\linewidth]{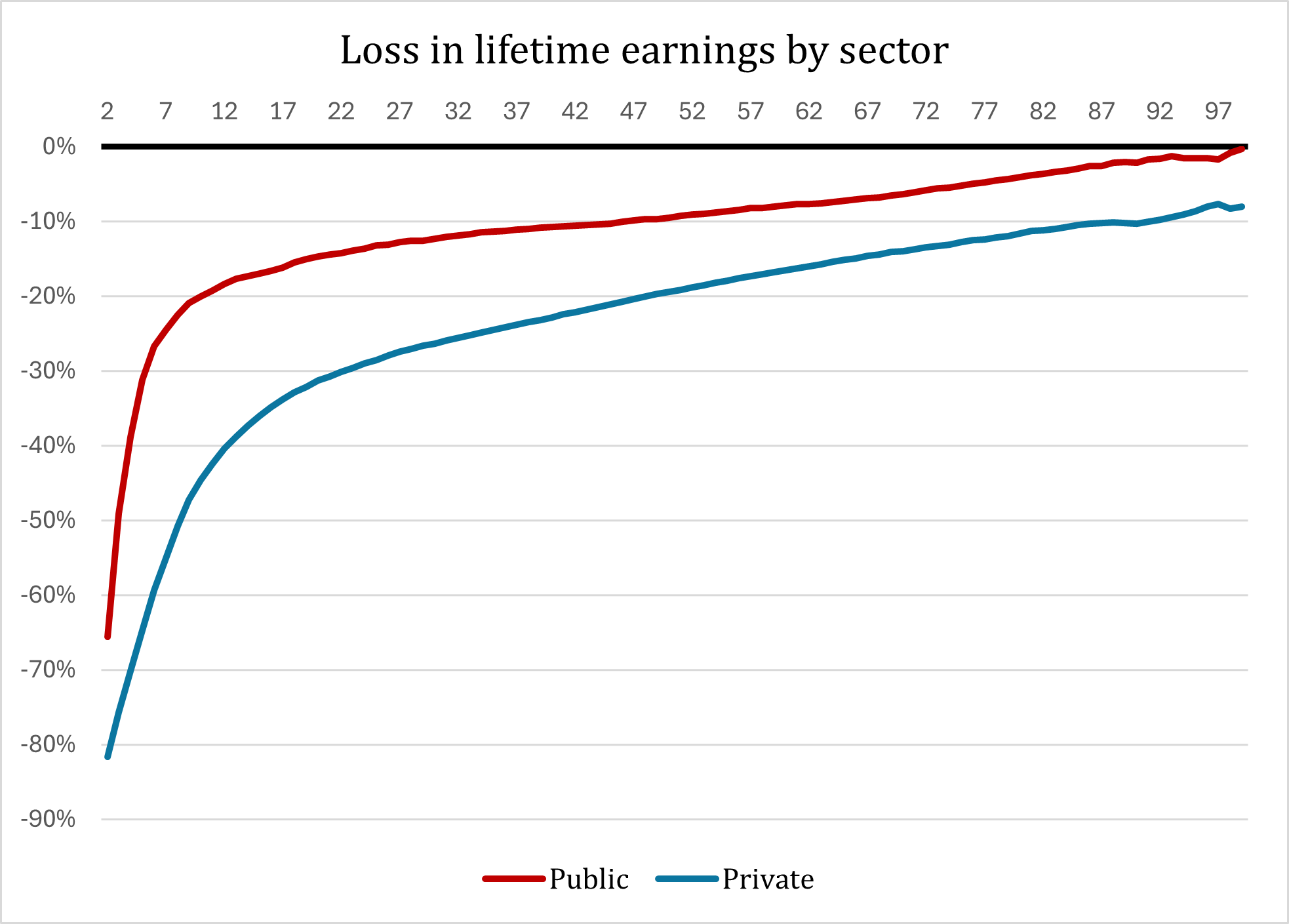}
        \caption{Loss in lifetime earnings with selection\\\scriptsize Lifetime earnings percentage difference of job-for-life case and mobility case (the latter takes into account the probability of transitioning to non-employment or part-time work).}
       
        \label{fig:enter-label}
    \end{figure}

\subsection{Men benefit from a gender premium in both sectors} \label{sec_gender_prem}
\begin{figure}
    \centering
    \includegraphics[width=0.7\linewidth]{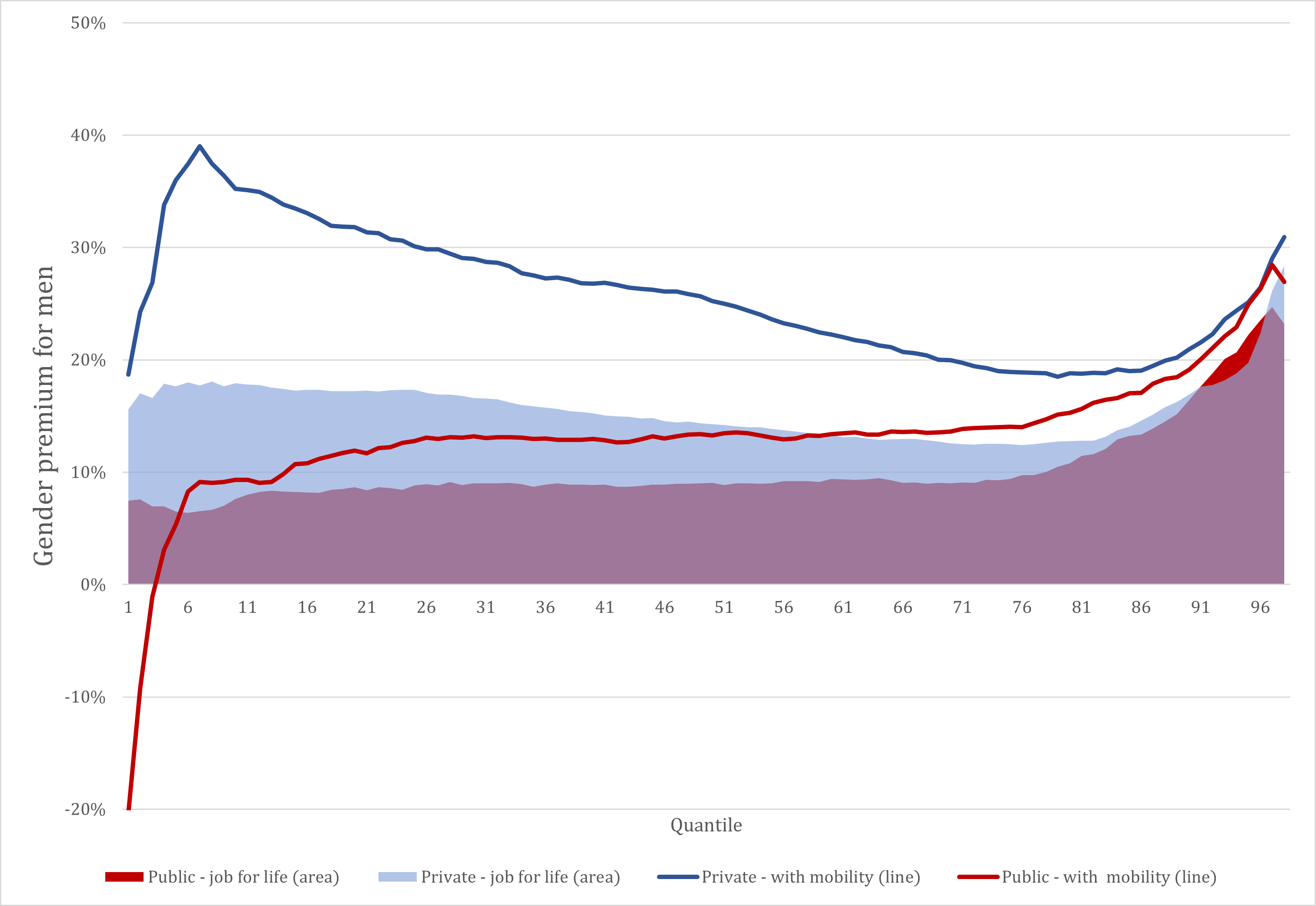}
    \caption{Gender premium in lifetime earnings}
    \label{fig:gender_prem}
\end{figure}

\noindent Whether considering the job-for-life case in counterfactual lifetime earnings or the case with mobility, the takeaway remains that men benefit from a gender premium in both sectors. This comparison can be made in two ways: given selection into working full-time in either sector, the lifetime earnings quantiles of men and women can be compared sectorally. Alternatively, their lifetime earnings with mobility can be compared, again conditional on selection. Figure \ref{fig:gender_prem} graphically depicts both these cases. As expected, the case that takes mobility into account (lines in the figure) produces higher gender gaps relative to the job for life case (area in the figure), particularly in the private sector. The public sector gender premium is almost always lower compared to private. A deeper dive into these differences by gender and by education categories reveals that on average, higher education lowers this gender premium that men benefit from in both sectors, even though glass ceilings exist in both sectors.

\noindent A particularly novel insight emerges when analyzing the distribution of lifetime earnings quantiles between men and women in each sector. The findings show that the male gender premium is present across the entire distribution but is more pronounced at higher earnings quantiles, especially in the private sector. This indicates that as earnings increase, men benefit from larger relative advantages, whereas women's wages remain comparatively compressed, particularly in the public sector. Although much of the literature on gender pay gaps focuses on penalties faced by women, this study provides new quantitative evidence on the persistence of male wage advantages across the earnings distribution. Even in a labor market where public employment offers greater stability and wage regulation, men continue to systematically earn more than women at respective lifetime earnings quantiles, reinforcing cumulative gender inequalities over the career cycle.

\subsection{Returns to education are lower in the public sector...}

\begin{figure}[H]
    \centering
    \includegraphics[width=0.9\linewidth]{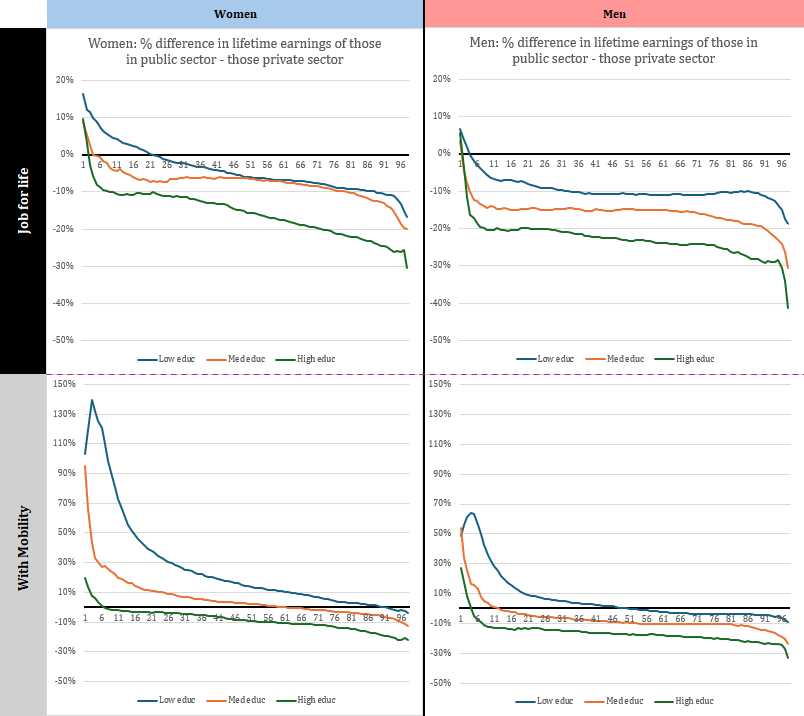}
    \caption{Percentage difference of lifetime earnings in the public and private sectors by education levels and sex}
    \label{fig:returs_educ}
\end{figure}

    \begin{itemize}
         \item \textbf{...for men and for women}
         Figure \ref{fig:returs_educ} illustrates that returns to education are consistently lower in the public sector compared to the private sector, for both men and women, across all lifetime earnings quantiles conditional on selection. The negative gap in lifetime earnings differences between public and private sector workers increases with education level, with highly educated workers experiencing the largest lifetime earnings penalty in the public sector.
        \begin{itemize}
            \item Among low-educated workers, differences between public and private sector earnings remain moderate, with some evidence of a public sector premium at the lowest quantiles. This suggests that for lower-income workers, public sector employment still offers some earnings security benefits, particularly in early career stages.
            \item However, for medium-educated workers, and even more so for highly educated workers, the public sector disadvantage becomes increasingly pronounced across the distribution, with lifetime earnings penalties widening at the upper quantiles. This suggests limited wage progression for highly skilled individuals in public sector jobs compared to their private sector counterparts.
            \item  The penalty for highly educated workers is particularly severe for men, reinforcing the idea that private sector careers provide greater returns to education through steeper earnings trajectories over a lifetime.
        \end{itemize}    
These findings confirm that while public employment provides stability, it does so at the cost of significantly lower lifetime earnings, especially for those with higher educational attainment.
         \item \textbf{...but job stability and transitions to public part-time reduces these penalties, particularly for women}
         Despite the lower returns to education in the public sector, women, particularly those in the lower earnings quantiles, experience significant lifetime earnings advantages in the public sector when accounting for career stability and part-time transitions.
         \begin{itemize}
             \item The "With Mobility" panels in Figure 9 show that women, particularly low-educated women, benefit significantly from public sector job stability. The public sector mitigates earnings volatility, allowing for a smoother and more predictable lifetime earnings trajectory.
             \item The gap in lifetime earnings between public and private employment narrows significantly for women once sectoral transitions are considered, suggesting that career interruptions and part-time work are better accommodated in the public sector. This is particularly relevant for women with medium and low education levels, who benefit from higher wage stability when transitioning to public part-time work.
         \end{itemize}
    \end{itemize}
These findings highlight a key gendered dynamic: while the private sector provides higher returns to education, the public sector acts as a critical stabilizing force for women, particularly in lower and middle-income categories. Public sector employment appears to buffer the income shocks associated with career breaks and transitions to part-time work, making it a more viable long-term option for many women, despite lower formal returns to education.



    \section{Discussion}\label{discussion_section}
    \noindent The findings of this study demonstrate that the model successfully captures observed income trajectories and sectoral mobility patterns, providing a robust framework for analyzing lifetime earnings differentials between public and private sector workers. Notably, the model replicates the substantial earnings dispersion between men and women, particularly in the private sector, where wage trajectories exhibit greater volatility and gender disparities are more pronounced. However, while the model effectively accounts for sectoral differences in wage dynamics and transitions, institutional factors beyond direct wage earnings—such as pension schemes—remain an important avenue for future research. The French public and private sector pension systems differ significantly, with public sector workers typically benefiting from more stable retirement incomes. While these differences are not explicitly captured in this dataset, policy changes in pension structures could have significant implications for lifetime earnings differentials across sectors. Future research could extend this analysis by incorporating expected retirement benefits and pension wealth accumulation to provide a more comprehensive measure of lifetime earnings.

\noindent Another promising avenue for future work lies in the integration of fertility and household employment decisions into the modeling framework. Labor market outcomes are often shaped by household-level choices, particularly for women, whose career trajectories are more likely to be influenced by caregiving responsibilities. A dynamic approach that accounts for joint household decision-making and the impact of parental leave policies could offer deeper insights into the gendered effects of public versus private sector employment.

\noindent Finally, the findings suggest broader implications beyond the French labor market. The gendered lifetime public premium observed in this study is likely to be even more pronounced in developing economies, where public sector employment often provides one of the few stable career pathways. However, the extent of this effect will depend on institutional labor protections, the degree of informality in private sector employment, and access to social security systems. Future comparative research could test this hypothesis by analyzing cross-country differences in public-private earnings trajectories and gender disparities, particularly in economies where public employment is a critical employer for educated women.

\noindent By expanding on these dimensions—pension structures, household decision-making, and cross-country comparisons—future research can further refine our understanding of how sectoral employment choices shape lifetime earnings and economic inequalities across gender and educational backgrounds.

    \section{Conclusion}\label{conclusion_section}
    \noindent This study reveals how France’s labor market institutions—centralized wage bargaining, strong job protections, and regulated part-time work—shape lifetime earnings disparities between public and private sectors. By modeling unobserved heterogeneity and career transitions, I reconcile conflicting prior findings: while hourly wage analyses suggest penalties for educated workers, lifetime earnings reveal a public premium for women and low-skilled employees, driven by stability and part-time compensation. Yet this premium comes at a cost. Highly educated workers trade a part of their lifetime earnings in the public sector, reflecting rigid wage progression that stifles returns to education.

    \noindent These results carry implications beyond France. As a prototype of coordinated market economies, France’s experience highlights the dual role of public sectors: mitigating gender inequality through stability while imposing ceilings on mobility. The results confirm the existence of a public premium in lifetime earnings, particularly for women, low-educated workers, and older employees. Even when accounting for selection effects, the public sector offers significant advantages to these groups in terms of income stability and reduced volatility. Policymakers in similar contexts (e.g., Scandinavia, Southern Europe) must weigh privatization’s risks—eroding gender equity gains—against the need to reform rigid wage structures. For developing economies, where public sectors often dominate formal employment, the findings underscore the potential of public jobs to reduce inequality.

    \noindent Future research should extend this framework to incorporate pension benefits and household dynamics, deepening our understanding of how institutions shape lifetime disparities. By bridging the gaps between literature on gender, education, and sectoral choice, this study offers a blueprint for analyzing equity-efficiency trade-offs in labor markets globally.

\bibliography{refs}

\appendix
\renewcommand{\thefigure}{A.\arabic{figure}}
\setcounter{figure}{0} 

\newpage
 \section{Appendix}\label{section::10}
    \subsection{Panel Construction} \label{panel_constr}
        \begin{itemize}
            \item Keep observations of people aged 18-60 and atleast 3 observed spells (including non-employment).
            \item Keep employment spells that are at least 6 months (180 days) in a year. Else, code as non-employment. Recall in DADS this employment durations include paid days/time off.
            \item Non-employment is imputed for the two following cases: 
                \begin{enumerate}
                    \item if someone has missing years in between employment spells. 
                    \item if someone who would be younger than 60 in 2019 vanishes from the panel.
                \end{enumerate}
                Every year from 2012-2019, about half of the non-employment spells have some non-employment compensation (chomage indemnisee).
            \item In constructing a panel that has one spell per individual per year, the highest paying spell was kept (note only employment spells of atleast 6 months are kept).
             \item Winsorize log net real annual earnings (variable \texttt{s\_net}) by state and year: cut and replace top and bottom 1\%.
            \item \textbf{Note:} Full-time hours in DADS is 1820 hours annually (this figure translates to a 35-hour workweek). This is because in DADS paid time off is also included in employment durations and in hours; so, 35-hour workweeks over the 52 weeks in a year gives us $35*52=1820$ hours annually. 
        \end{itemize}

        \begin{table}[htbp]\centering
                        \def\sym#1{\ifmmode^{#1}\else\(^{#1}\)\fi}
                        \caption{Classification of French Degrees and U.S. Equivalents\label{tab::educ}}
                        \scriptsize 
                        \begin{tabular*}{1\hsize}{p{0.1\linewidth} | p{0.3\linewidth} | p{0.3\linewidth} | p{0.2\linewidth}}
                        \toprule
                                            Category&Degree&U.S. Equivalent&Sample Classification\\ \hline
                        \midrule
                        1                &     Aucun Diplôme  déclaré      &     No Degree declared        &             \\
                        2 &      CEP, DFEO         &      Elementary or Middle School         &              \\
                        3            &       BEPC, BE, BEPS  &       High School  &         \\
                        4&      CAP, BEP, EFAA, BAA, BPA, FPA 1er      & Vocational Technical School (Basic) &                      \\
                        5& Baccalauréats technique et professionnel, brevet professionnel, autres brevets (BEA, BEC, BEH, BEI, BES, BATA), baccalauréat général, brevet supérieur, CFES & High school diploma (General or Technical/Vocational) &      Low \\
                        \hline
                        6& Santé, BTS, DUT, DEST, DEUL, DEUS, DEUG &    Associate's degree or equivalent vocational degree &      Medium         \\
                        \hline
                        7& 2ème cycle, 3ème cycle, Grande école, CAPES, CAPET &  Graduate School and Other Post-Secondary Education&      High         \\
                        \bottomrule
                        \end{tabular*}
                    \end{table}
    \subsection{Descriptives}
    
     	\subsubsection*{Public Share of Employment}\label{section::3.3}
	We observe that the public share of total employment in the sample remains fairly stable with an overall gradual increase of approximately 1.4\% rising from a little under 23\% close to 24.6\% over the 8-year period. The slight but stable increase in the public share of employment is in-line with official INSEE statistics and corroborated in this sample panel.
    \subsubsection{Additional demographics in the Public and Private Sectors}
        		\begin{figure}[htbp]
                      \centering
                      \caption{Share of women in the Public and Private Sectors over 2012-19}
                      \includegraphics[width=10cm]{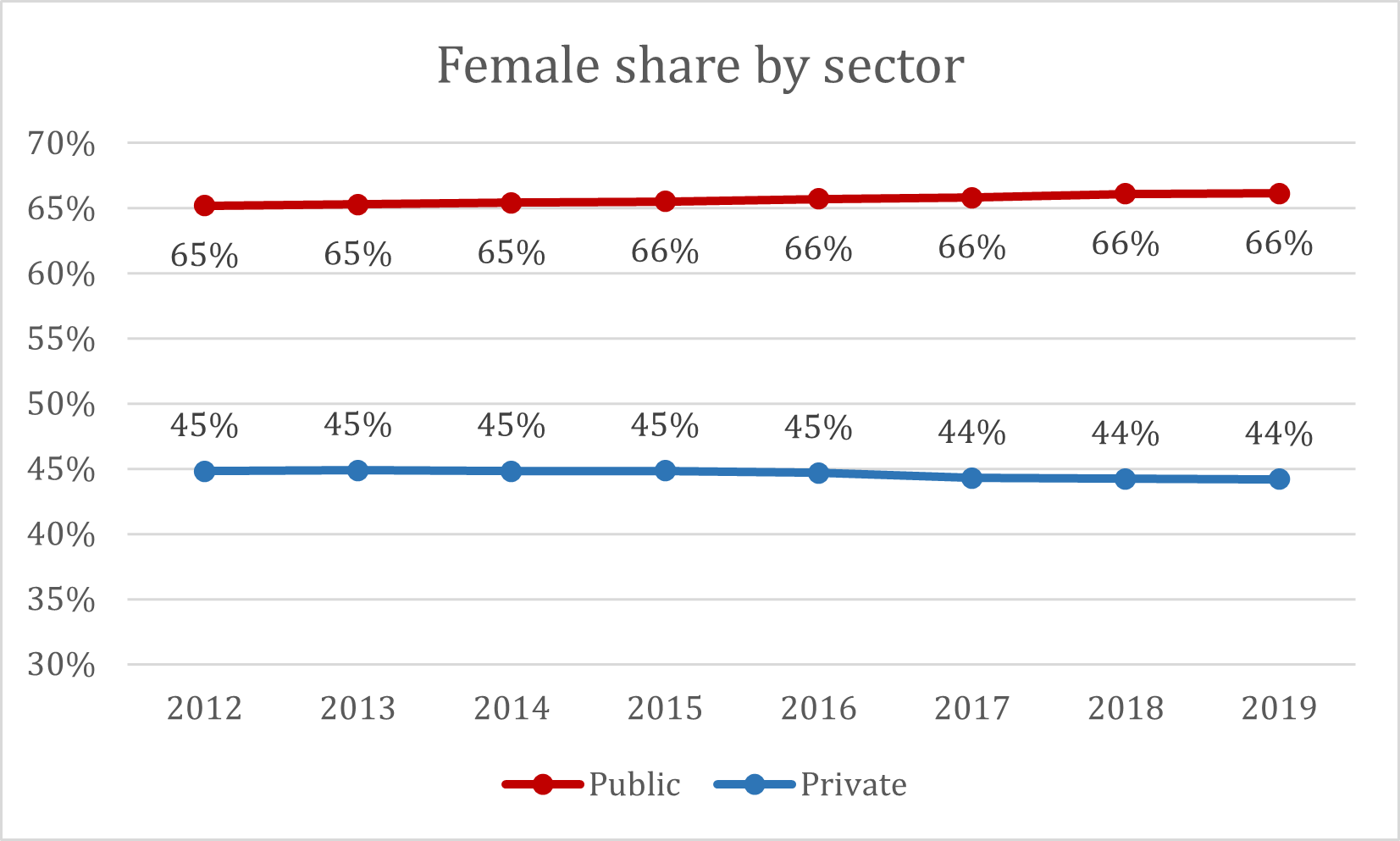}
                      \label{fig::sexratiopubpvt}
                \end{figure}
                The share of employment in both sectors is stable across the years observed. The public sector, including part-time and full-time work averages at 24.3\%, with small deviations. Since this article focuses on the intersection with gender, the sex ratio in both sectors, by full-time status is of key importance. Women hold 60\% of the full-time public sector jobs in contrast to the 39\% in the private sector. Interestingly, the part-time work landscape is very different as most of these jobs are done by women. Over the years, the female share of part-time jobs rises in both sectors . 
                
                The stability of the sex ratio (of females to males) in the public and private sectors irrespective of part-time full-time worker status is captured by Figure \ref{fig::sexratiopubpvt} generated from the sample. In France, the sex ratio is roughly 65:35 in the public sector and 45:55 in private. Curiously, women's earnings in the public sector is more compressed than that of men and also peaks later for the mean and above-mean income earners. The picture in the private sector by sex is even more grim. The dispersion of women's earnings far exceeds that of men. The lowest decile earners across sex experience a significant fall in earnings towards retirement. However, the fall not only starts relatively earlier in women's trajectories, but is far steeper across income percentiles. The below table depicts which sector, gender, and age group all the individuals observed in the panel (using their first spell) belong to:
            \begin{table}[h]
            \centering
            \caption{Distribution of Education Levels by Sector, Sex and Age Groups at first observed spell}
            \scriptsize 
            \begin{tabular}{llcccccc}
            \toprule
            \multirow{2}{*}{Sex} & \multirow{2}{*}{Education} & \multicolumn{3}{c}{Public Sector} & \multicolumn{3}{c}{Private Sector} \\
            \cmidrule(lr){3-5}\cmidrule(lr){6-8}
            & & Under 30 & 31-45 & Over 45 & Under 30 & 31-45 & Over 45 \\
            \midrule
            \multirow{3}{*}{Females} & Low & 15 & 36 & 49 & 23 & 37 & 40 \\
            & Medium & 19 & 48 &  34& 27 & 51 & 22 \\
            & High & 17 & 58 & 26 & 31 & 52 & 17 \\
            \midrule
            \multirow{3}{*}{Males} & Low & 15 & 38 & 46 & 26 & 39 & 35 \\
            & Medium & 15 & 48 & 37 & 26 & 51 &  23\\
            & High & 12 & 53 & 35 & 27 & 50 & 23 \\
            \bottomrule
            \end{tabular}
            \end{table}

            \begin{figure}
                \centering
                \includegraphics[width=0.7\linewidth]{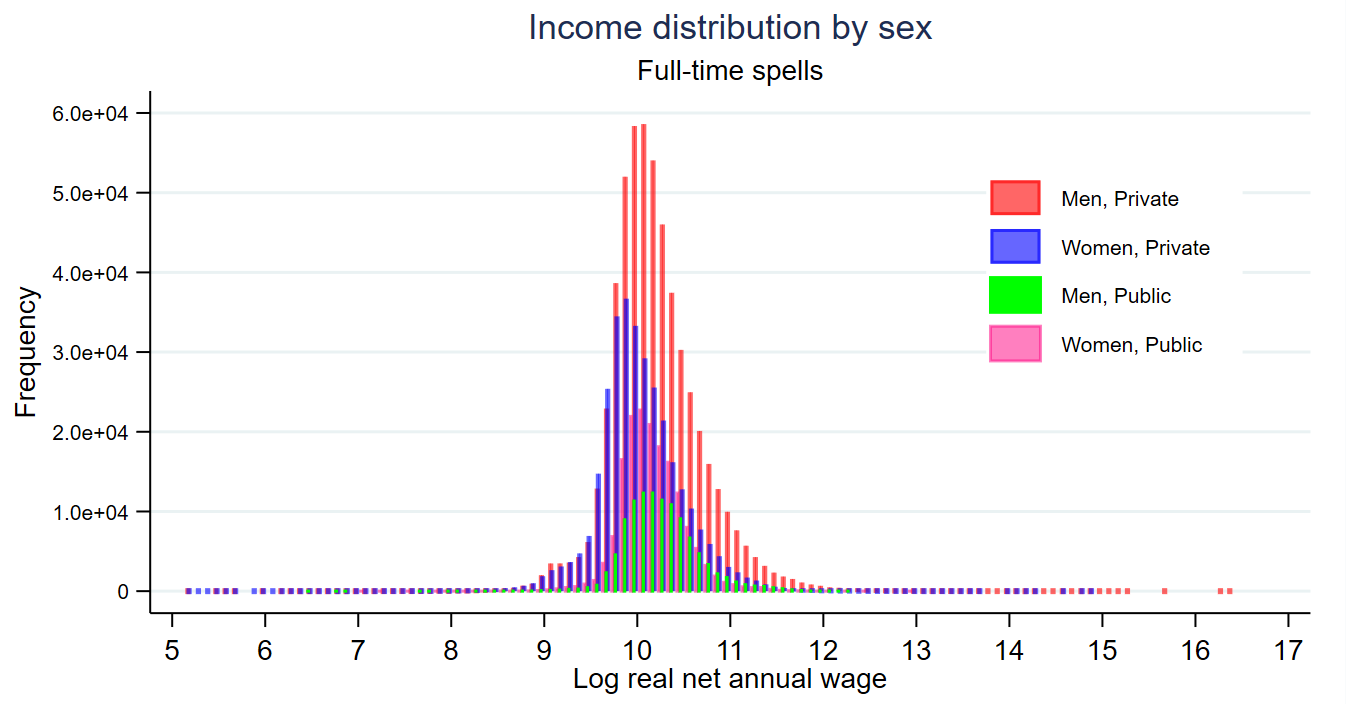}
                \caption{}
                \label{fig::stud}
            \end{figure}

        \subsubsection*{Income and non-employment Statistics}\label{section::3.4}
        
    The non-employment or inactivity remained mostly stable during this period at about 30\%. Since it is imputed from the dataset it underestimates the rate for the first two years but catches up thereafter. Figure \ref{fig::stud} shows the income distribution of men and women in the public and private sectors over the whole sample. Recall that this includes part-time workers. There are many noteworthy points in this distribution : firstly, the dominance of males in the private sector and of females in the public sector is evident. Secondly, as expected, the spread of income is more dispersed in the private sector while it is relatively more compressed in the public sector. Thirdly, the private sector distribution of income of males to the right of the mean dominates that of women; implying that a higher number of men earn more than the average than do women: notice the blue bars to the right of the mean in the figure. This phenomenon is reversed in the public sector distributions where more women earn the mean and above mean incomes. But, since the public sector is relatively smaller than the private, the public distributions are much shorter than their private counterparts. This income distribution provides initial evidence for the average earnings of females being lower than for men in the private sector. 

\begin{table}[H]
\centering
\tiny
\tiny \caption{OLS regressions on real net wages - alternative specification}
\label{tab:regression_results_extra}
\begin{tabular}{lcc}
\hline
VARIABLES & Net annual & Net hourly \\
\hline
Pub FT & 0.100*** & -0.006 \\
 & (0.005) & (0.004) \\
Pvt PT & -0.636*** & -0.118*** \\
 & (0.018) & (0.010) \\
Pub PT & -0.443*** & -0.097*** \\
 & (0.009) & (0.005) \\
\textcolor{red}{Female} & \color{red}{-0.238***} & \color{red}{-0.171***} \\
 & (0.011) & (0.008) \\
\color{red}{Pub FT\#1.female} & \color{red}{0.117***} & \color{red}{0.059***} \\
 & (0.011) & (0.008) \\
 Pvt PT\#1.female & 0.117*** & 0.071*** \\
 & (0.028) & (0.016) \\
 Pub PT\#1.female & 0.337*** & 0.121*** \\
 & (0.013) & (0.009) \\
Experience (over a decade) & 0.374*** & 0.275*** \\
 & (0.002) & (0.001) \\
Experience over a decade squared & -0.065*** & -0.042*** \\
 & (0.000) & (0.000) \\
Medium educ & 0.288*** & 0.264*** \\
 & (0.001) & (0.001) \\
High educ & 0.537*** & 0.529*** \\
 & (0.001) & (0.001) \\
Constant & 9.498*** & 2.229*** \\
 & (0.006) & (0.004) \\
\hline
Observations & 1,389,551 & 1,385,227 \\
Adjusted R-squared & 0.454 & 0.387 \\
Year fixed effects & \checkmark & \checkmark \\
Contract x state x female fixed effects & \checkmark & \checkmark \\
\hline
\multicolumn{3}{l}{Standard errors in parentheses} \\
\multicolumn{3}{l}{*** p<0.01, ** p<0.05, * p<0.1} \\
\end{tabular}
\end{table}

  \subsection{Model} \label{app_model}
        This subsection lays out the equations that are similar in spirit to \citet{ppgap} and \citet{ppgapeur} - the only difference being the data structure and the employment $state$ variable.
    \subsubsection{Unobserved Heterogeneity} \label{app_unob_het}
    
The probability of an individual belonging in a particular class given their observed individual heterogeneity $z_i^f$ is given as 
 \begin{equation}\ \label{kprob}
    \Pr\{k_i^y,k_i^m|z_i^f\}=\Pr\{k_i^y|k_i^m,z_i^f\}\cdot \Pr\{k_i^m|z_i^f\}
\end{equation}
In this analysis, given four employment states, 4 transition classes \& 3 income classes are settled upon after trying various other combinations, which give the best compromise between analytic efficiency and accuracy of model fit. Note that both $k^m$ and $k^y$ are categorical variables. Both components of \eqref{kprob} are expressed as multinomial logits with $K^y=3$ and $K^m=4$ outcomes respectively. Based on a standard multinomial logit classification, the functional form of the probabilities of an individual belonging to these latent classes can be described as follows:
\begin{equation} \label{kcoeff}
    \begin{split}
        \Pr \{k_i^m=k^m \mid z_i^f\} = \frac{\exp\left(z_i^{f\prime} \kappa^m_{k^m}\right)}{\sum_{k=1}^{K^{m}} \exp\left(z_i^{f\prime} \kappa^m_k\right)}, \\
        \Pr \{k_i^y=k^y \mid k_i^m, z_i^f\} = \frac{\exp\left( \begin{psmallmatrix} z_i^f \\ k_i^m \end{psmallmatrix}^{\!\prime} \kappa^y_{k^y} \right)}{\sum_{k=1}^{K^{y}} \exp\left( \begin{psmallmatrix} z_i^f \\ k_i^m \end{psmallmatrix}^{\!\prime} \kappa^y_k \right)},
    \end{split}
\end{equation}

\noindent where $k_i^m$ takes the values 1, 2, 3 or 4 for an individual $i$. Similarly, $k_i^y$ takes the values 1, 2 or 3 for an individual $i$. Here, $\kappa_k^m$ and $\kappa_k^y$ are the vectors of coefficients of respective multinomial logits for the $k^{th}$ outcome. Taking 1 as the base outcome in each multinomial logit, $\kappa_1^m$ and $\kappa_1^y$ are normalized to zero.

\subsubsection{Transitions in labor market states and Transition classes: \texorpdfstring{$K^m$}{Km}}\label{section::5.3}
Moving on to the second component of the individual likelihood equation \eqref{lhood}, which encompasses the information in the model relating to the transitions between the various labor market states. Labor market states at year-t are assumed to depend on the individual's previous labor market state and their observed and unobserved heterogeneity, i.e., it follows a conditional first-order Markov chain. As there are five possible states, the probability of the individual being in one of the five states is modeled as a multinomial logit where the explanatory variables are the previous labor market state and the individual's unobserved and observed heterogeneity. Thus, at any given year (t > 1), 
 \begin{equation}\ \label{chcoeff}
    \Pr \{S_{i,t}=S_s | S_{i,t-1}, z_{i,t-1}^v, z_i^f, k_i^m\}=\ddfrac{\exp\Bigg( \begin{psmallmatrix} S_{i,t-1}\\ z_{i,t-1}^v\\ z_i^f\\k_i^m\end{psmallmatrix}'  \chi_{S_s}\Bigg)}{\sum_{S=0}^4 \exp\Bigg(\begin{psmallmatrix}S_{i,t-1}\\ z_{i,t-1}^v\\ z_i^f\\k_i^m\end{psmallmatrix}'  \chi_S\Bigg)}
\end{equation}
This equation is valid for all years except for the first year as the first state is the initial value. The initial state is specified as being dependent only on the observed and unobserved heterogeneity and thus, similarly modeled as a multinomial logit:
 \begin{equation}\ \label{chicoeff}
    \Pr \{S_{i,1}=S_s | z_i^f, k_i^m\}=\ddfrac{\exp \bigg( \begin{psmallmatrix} z_i^f\\k_i^m\end{psmallmatrix}'  \chi_{S_s}^o\bigg)}{\sum_{S=0}^4 \exp \bigg(\begin{psmallmatrix}z_i^f\\k_i^m\end{psmallmatrix}'  \chi_S^o\bigg)}
\end{equation}
$[\chi_s]_{s=0}^{S-1}$ and $[\chi_s^o]_{s=0}^{S-1}$ are the vectors of coefficients of the respective multinomial logits for the $S^{th}$ state outcome. $\chi_0^{}$ and $\chi_0^o$ are the base outcomes and normalized at zero.

Given these two specifications, we can now construct the (conditional) likelihood function of observing an individual's employment state trajectory over the years as follows:  

 \begin{equation}\label{lhoodm}
    \ell_i(S_{it}|z_{it}^v, z_i^f,k_i^m) =  \Pr\{S_{i1}|z_i^f,k_i^m\} \times \prod_{t=2}^{T_i} \Pr\{S_{it} | S_{i,t-1},z_{i,t-1}^v, z_i^f,k_i^m \}  
\end{equation}

\subsection{Estimating the mobility and income parameters using sequential EM}\label{section::5.5}
    A sequential, limited information version of the EM algorithm inspired by \citet{br1} is employed to obtain a consistent set of estimates. In the previous sections, one can see all the parameters of the model that need to be estimated. The entire list of parameters can be split into two:
    $\Theta^m=\big\{ (\kappa_k^m)^{K^m-1}_{k=0}, (\chi_s)^{4}_{s=0}, (\chi^0_{s})^{4}_{s=0} \big\}$ and  $\Theta^y=\big\{ (\kappa_k^y)^{K^y-1}_{k=0}, \mu, \sigma, \xi \big\}$. To recollect, we get the $\Theta^m$ parameters from the state mobility equations \eqref{kcoeff}, \eqref{chcoeff} and \eqref{chicoeff}; and we get the $\Theta^y$ parameters from the income process equations \eqref{kcoeff}, \eqref{mu}, \eqref{sigma}, \eqref{tau1}.
    The structure of the complete individual likelihood equation \eqref{lhood} means that it can be decomposed into two parts as shown below:
        \begin{gather} \label{compilhood}
             \mathcal{L}_i(\mathbf{x}_i, k_i; \Theta^m,\Theta^y) = \mathcal{L}_i^m(\mathbf{x}_i, k_i^m; \Theta^m) \times \mathcal{L}_i^y(\mathbf{x}_i, k_i^m, k_i^y; \Theta^y) \\ \nonumber
             \\ \text{where }\nonumber   \mathcal{L}_i^m(\mathbf{x}_i,k_i^m; \Theta^m) = \ell_i^m \Big(\mathbf{S}_i | \mathbf{z}_i^v, z_i^f, k_i^m; \Theta^m  \Big) \cdot \Pr \Big\{ k_i^m | z_i^f ; \Theta^m \Big\} \text{, }\\
             \nonumber \text{and } \mathcal{L}_i^y(\mathbf{x}_i,k_i^m,k_i^y; \Theta^y) = \ell_i^y \Big(\mathbf{y}_i | S_i, \mathbf{z}_i^v, z_i^f, k_i^m,k_i^y; \Theta^y  \Big) \cdot \Pr \Big\{ k_i^y | k_i^y, z_i^f ; \Theta^m \Big\}
        \end{gather}
This structure makes it easier to separate income sequences $(\mathbf{y_i})$ and income classes ($k_i^y$) from the $state$ transition part of the likelihood function, $ \mathcal{L}_i^m(\mathbf{x}_i,k_i^m; \Theta^m)$. The parameters pertaining to the job mobility process and job transition classes can thus be recovered by separately considering the likelihood of observed job sector mobility: $\sum_{i=1}^{N} \ln \Big(\sum_{k^m_i=1}^{K^m}  \mathcal{L}_i^{m} (\mathbf{x}_i,k_i^m; \Theta^m) \Big)$. The maximization can be achieved by applying the EM algorithm for finite mixtures. To make the process more efficient, the parameters for the $state$ transitions part of the likelihood equation are first estimated. This yields an initial estimate of $\hat{\Theta}^m$ for the mobility parameters. Then, $\Theta^m$ is fixed at $\hat{\Theta}^m$, and the estimation proceeds to the income parameters $\Theta^y$. After reaching a suitable convergence in terms of distance between the parameters from subsequent iterations, an initial estimate $\hat{\Theta}^y$ of $\Theta^y$ is obtained. Finally, using $\hat{\Theta}^m$ and $\hat{\Theta}^y$ as initial values, the complete likelihood function (given by equation \eqref{lhood} and \eqref{compilhood}) is estimated, allowing all parameters to vary, to arrive at the final estimates of all the mobility and income parameters.

\subsubsection{Calibrating the initial job mobility parameters \texorpdfstring{$\Theta^m$}{Thetam}}\label{section::5.5.1}

\begin{itemize}
    \item \textbf{E-Step}: For an initial value $\Theta^m_n$ of $\Theta^m$, for each transition class $k^m = 1,...,K^m$ and for each individual $i$ in the sample, the posterior probability that $i$ belongs to transition class $k^m$ given $\mathbf{x}_i$ and $\Theta^m_n$ would be:
     \begin{equation}\label{kmprob}
        \Pr \{k_i^m = k^m | \mathbf{x}_i; \Theta^m_n \} = \ddfrac{\mathcal{L}_i^{m} (\mathbf{x}_i,k_i^m; \Theta^m_n)}{\sum_{k=1}^{K^m}  \mathcal{L}_i^{m} (\mathbf{x}_i,k_i^m; \Theta^m_n)}
        \end{equation}
    These probabilities are calculated by estimating the likelihoods using using the functional forms of equations \eqref{kcoeff}, \eqref{chcoeff} and \eqref{chicoeff}.  

    \item \textbf{M-Step}: We now update $\Theta_n^m$ into $\Theta_{n+1}^m$ by maximizing the following augmented sample log-likelihood, weighted by the E-Step equation above:
     \begin{equation}\label{mmax}
        \Theta_{n+1}^m = \argmax_{\Theta^m}  \sum_{i=1}^{N} \sum_{k_i^m=1}^{K^m} \Pr\Big\{ k_i^m = k | \mathbf{x}_i; \Theta_n^m\Big\} \cdot \ln \Big[ \mathcal{L}_i^{m} (\mathbf{x}_i,k_i^m; \Theta^m_n) \Big]
    \end{equation}
    This maximization is carried out by creating 4 copies of out data set, one for each possible transition class that an individual can be in. Then we carry out weighted multinomial logits of $k^m$, $State$, and $State_{initial}$ using equations \eqref{kcoeff}, \eqref{chcoeff} and \eqref{chicoeff} weighing each copy of the data by its prior probability obtained in the E-step for the corresponding transition class $k^m$.     
\end{itemize}
We continue iterations till the Euclidean distance between the coefficients in subsequent iterations falls below $10^{-3}$ as is standard in the literature. We can now use this initial estimate $\hat{\Theta}^m$ to calibrate the income process estimations in the next step. 
\subsubsection{Calibrating the income parameters \texorpdfstring{$\Theta^y$}{Thetay}}\label{section::5.5.2}
\begin{itemize}
    \item \textbf{E-Step}: For an initial value $\Theta_n^y$ of $\Theta^y$, for each class index $k = (k^m, k^y)$, $k^m = 1, . . . ,K^m; k^y = 1, . . . ,K^y$, and for each individual $i$ in the sample, compute the posterior probability that $i$ belongs to transition class $k^m$ and income class $k^y$ given $\mathbf{x}_i$, $\Theta_n^y$ and $\Hat{\Theta}^m$:
     \begin{equation}\label{ymprob}
        \Pr \Big\{ k_i^m = k^m, k_i^y = k^y | \mathbf{x}_i; \Hat{\Theta}^m, \Theta_n^y \Big\} = \ddfrac{\mathcal{L}_i \Big( \mathbf{x}_i, k^m, k^y; \Hat{\Theta}^m,  \Theta_n^y  \Big)}{\sum_{\ell^m=1}^{K^m} \sum_{\ell^y=1}^{K^y} \mathcal{L}_i \Big( \mathbf{x}_i, \ell^m, \ell^y; \Hat{\Theta}^m,  \Theta_n^y  \Big)}
    \end{equation}
    \item \textbf{M-Step}: In this step we update the $\mu, \sigma,  \xi \text{ and } \kappa^y $ coefficients. The following process is followed:
    \begin{enumerate}
        \item We first create 12 copies of our dataset. One for each $k^y$ and $k^m$ class combination that every individual can be in. Each class combination is weighted by the class probabilities obtained from equation \eqref{ymprob}.
        \item We then update the income mean parameter $\mu(\cdot)$ using weighted OLS regressions of $y_{it}$ on $\Big( S_{it},z_{it}^v,z_i^f,k_i^y\Big)$, using the 12 class probabilities obtained from equation \eqref{ymprob} in the E-Step as weights for each of the 12 copies of the data. The vector of coefficients of this regression thus becomes the updated $\hat{\mu}^{n+1}(\cdot)$.
        \item Similarly, the log squared residuals from the regression in step 2 are taken and regressed on $\Big( S_{it},z_{it}^v,z_i^f,k_i^y \Big)$, again using weighted OLS, to update income variance parameter $\hat{\sigma}^{n+1}(\cdot)$.
        \item Log income disturbances are updated as:
        \begin{gather}
            \nonumber \Tilde{y}_{it}^{n+1} = \frac{y_{it} - \hat{\mu}_{n+1}(S_{it},z_{it}^v,z_{i}^f,k_i^y)}{\hat{\sigma}_{n+1}(S_{it},z_{it}^v,z_{i}^f,k_i^y)}
        \end{gather}
        Next, we can generate the $\tau_{i,t,t-1}$ of $\Tilde{y}_{it}$ by taking into account all past income streams up until the year $t$. So, for each year, the $Cov ( \Tilde{y}_{it}^{n+1}, \Tilde{y}_{i,t-1}^{n+1})$ measure is expected to be the outcome of the $\tau_1^{n+1}(\cdot)$ function. Thus, we can update $\hat{\tau}_1^{n+1}(\cdot)$ by taking the inverse functional form of $\tau_1^{n+1}$ that was given in \eqref{tau1}. To do this, a new variable $f_{it}$ is defined such that:
        \[
        f_{it} = ln \Bigg( \ddfrac{1+Cov ( \Tilde{y}_{it}^{n+1}, \Tilde{y}_{i,t-1}^{n+1})}{1-Cov ( \Tilde{y}_{it}^{n+1}, \Tilde{y}_{i,t-1}^{n+1})} \Bigg)
        \]
       
       We then regress $f_{it}$ on $S_{it} , S_{i,t-1} , z_{it}^v ,k_i^m, k_i^y , S_{it}*k_i^y , S_{i,t-1}*k_i^y$, again using equation \eqref{ymprob} as weights. The vector of coefficients thus obtained is the updated $\xi^{n+1}$.
      
        
        \item Finally we can update the set of income class assignment parameters, $\big( \kappa^y_k\big)^{K^y-1}_{k=0}$ by running a weighted multinomial logit regression of class indices on $\big( z_i^f, k_i^m\big)$, again by using the \eqref{ymprob} as weights.
    \end{enumerate}
\end{itemize}
\subsubsection{Likelihood Maximization}\label{section::5.5.3}
After obtaining the initial estimates of the $\Theta^m$ and $\Theta^y$ parameters separately, we then proceed to update the combined log-likelihood given by the equation:
 \begin{gather}
     \text{likelihood: }\prod_{i=1}^{N}  \prod_{k_i^m=1}^{K^m} \prod_{k_i^y=1}^{K^y} \mathcal{L}_i \Big[\mathbf{x}_i, (k_i^m, k_i^y) \Big]\\
     \text{log-likelihood: }\sum_{i=1}^{N} \ln \Bigg( \sum_{k_i^m=1}^{K^m} \sum_{k_i^y=1}^{K^y} \mathcal{L}_i \Big[\mathbf{x}_i, (k_i^m, k_i^y) \Big] \Bigg)
 \end{gather}
We carry out this estimation using a similar methodology employed in the initialization phase.  

\subsection{Model Fit}
\subsubsection{Additional Tables and Figures for Model Fit and Unobserved Heterogeneity}	
\begin{table}[ht]
\centering
\scriptsize
\caption{Demographic and Educational Composition by Transition Class ($km$)}
\begin{tabular}{lccccccc}
\toprule
Transition & Female & Low ed & Med ed & High ed & Age $<$31 & Age 31--45 & Age $>$45 \\
\midrule
$km=0$ & 0.65 & 0.51 & 0.16 & 0.32 & 0.25 & 0.43 & 0.32 \\
$km=1$ & 0.66 & 0.71 & 0.29 & 0.00 & 0.32 & 0.40 & 0.28 \\
$km=2$ & 0.33 & 0.86 & 0.14 & 0.00 & 0.32 & 0.39 & 0.29 \\
$km=3$ & 0.56 & 0.68 & 0.10 & 0.22 & 0.23 & 0.41 & 0.36 \\
\bottomrule
\end{tabular}
\caption*{\textit{Note:} Means by class from the latent model output. Classes differ markedly in observed characteristics—e.g., gender and education—but overlap substantially across those dimensions, suggesting compositional clustering rather than latent behavioral types.}
\label{tab:km_composition}
\end{table}
\begin{table}[H]
    \centering
    \scriptsize
    \setlength{\tabcolsep}{4pt}
    \renewcommand{\arraystretch}{1.1}
    \caption{Transition Probabilities by Latent Transition Class ($km$) -- Observed Data}
    \begin{tabular}{@{}lccccc@{}}
    \toprule
    \textbf{From/To}     & NE & Pvt FT & Pub FT & Pvt PT & Pub PT \\
    \midrule
    \multicolumn{6}{c}{\textbf{Transition Class 0}} \\
    \midrule
    NE       & 0.81 & 0.10 & 0.02 & 0.05 & 0.01 \\
    Pvt FT& 0.08 & 0.89 & 0.00 & 0.03 & 0.00 \\
    Pub FT & 0.02 & 0.01 & 0.93 & 0.00 & 0.04 \\
   Pvt PT& 0.15 & 0.12 & 0.00 & 0.72 & 0.01 \\
    Pub PT & 0.08 & 0.01 & 0.14 & 0.01 & 0.76 \\
    \midrule
    \multicolumn{6}{c}{\textbf{Transition Class 1}} \\
    \midrule
    NE       & 0.82 & 0.10 & 0.02 & 0.05 & 0.01 \\
    Pvt FT   & 0.09 & 0.88 & 0.00 & 0.03 & 0.00 \\
    Pub FT   & 0.03 & 0.01 & 0.92 & 0.00 & 0.05 \\
    Pvt PT   & 0.17 & 0.11 & 0.00 & 0.71 & 0.00 \\
    Pub PT   & 0.10 & 0.01 & 0.11 & 0.01 & 0.76 \\
    \midrule
    \multicolumn{6}{c}{\textbf{Transition Class 2}} \\
    \midrule
    NE       & 0.83 & 0.11 & 0.01 & 0.04 & 0.01 \\
    Pvt FT   & 0.09 & 0.89 & 0.00 & 0.02 & 0.00 \\
    Pub FT   & 0.03 & 0.01 & 0.93 & 0.00 & 0.03 \\
    Pvt PT   & 0.19 & 0.13 & 0.00 & 0.67 & 0.00 \\
    Pub PT   & 0.12 & 0.01 & 0.13 & 0.01 & 0.72 \\
    \midrule
    \multicolumn{6}{c}{\textbf{Transition Class 3}} \\
    \midrule
    NE       & 0.82 & 0.10 & 0.02 & 0.05 & 0.01 \\
    Pvt FT   & 0.08 & 0.89 & 0.00 & 0.03 & 0.00 \\
    Pub FT   & 0.02 & 0.00 & 0.93 & 0.00 & 0.04 \\
    Pvt PT   & 0.17 & 0.11 & 0.00 & 0.71 & 0.01 \\
    Pub PT   & 0.09 & 0.01 & 0.13 & 0.01 & 0.76 \\
    \bottomrule
    \end{tabular}
    \vspace{0.5em}

    \caption*{\textit{Note:} Entries show the probability of remaining or moving between states from period $t$ to $t+1$,
    where NE = non-employment, Pvt FT = private full-time, Pub FT = public full-time, Pvt PT = private part-time, and Pub PT = public part-time.
    Transition probabilities across the four latent transition classes ($km=0$--$3$) show extremely similar magnitudes.
    The persistence in key employment states (0.81–0.83 for non-employment, 0.88–0.89 for private full-time, and 0.92–0.93 for public full-time) varies by less than one percentage point across classes.
    This similarity implies that the unobserved heterogeneity component ($km$) primarily reflects compositional differences in demographics and sectoral attachment rather than distinct behavioral adjustment regimes.}
    \label{tab_trans_km_observed}
\end{table}

  \begin{table}[H]
            \centering
            \footnotesize
            \setlength{\tabcolsep}{4pt}
             \linespread{1}\selectfont
            \caption{Summary of Latent Classes}
            \begin{tabular}
            {|p{0.06\textwidth}|p{0.06\textwidth}|p{0.06\textwidth}|p{0.06\textwidth}|
            p{0.06\textwidth}|p{0.1\textwidth}|p{0.06\textwidth}|p{0.12\textwidth}|p{0.25\textwidth}|}
            \hline
            \textbf{Latent Class} & \textbf{$k^y$}& \textbf{$k^m$} & \textbf{Class Size (\%)} & \textbf{Female Share} & \textbf{Age Group} & \textbf{Educ. Level} & \textbf{Income Mobility} & \textbf{Labor Market Dynamics} \\
            \hline
            \textbf{1} & $1$ & $1$ & 3.9\% & 49.3\% & 51.9\% over 45 & Mixed & Upward (Q5: 35.9\%) & High non-employment, mixed gender stability in Pvt FT, upward income mobility (Q5: 35.9\%) \\
            \hline
            \textbf{2} & $1$ & $2$   & 0.1\% & 50.9\% & 54.7\% over 45 & Low & Moderate (Q5: 20.4\%) & High non-employment, women slower to transition to Pvt FT, moderate upward mobility (Q5: 20.4\%) \\
            \hline
            \textbf{3} & $1$ & $3$   & 2.0\% & 20.8\% & 53.1\% over 45 & Low & Moderate (Q5: 22.4\%) & High non-employment, men stable in Pvt FT, moderate upward mobility (Q5: 22.4\%) \\
            \hline
            \textbf{4} & $1$ & $4$   & 2.0\% & 41.7\% & Balanced & Mixed & Upward (Q5: 30.2\%) & Balanced employment, more transitions to Pvt FT, upward mobility (Q5: 30.2\%) \\
            \hline
            \textbf{5} & $2$ & $1$ & 13.8\% & 71.1\% & 42.7\% over 45 & Low-Mid & Moderate (Q5: 23.9\%) & Diverse transitions, more women shifting to Pvt PT, moderate income mobility (Q5: 23.9\%) \\
            \hline
            \textbf{6} & $2$ & $2$  & 1.8\% & 68.6\% & Balanced & Low & Limited (Q5: 15.3\%) & High non-employment, women remain in Pvt PT longer, limited upward mobility (Q5: 15.3\%) \\
            \hline
            \textbf{7} & $2$ & $3$  & 12.4\% & 36.7\% & 41.5\% over 45 & Low & Moderate (Q5: 18.2\%) & Men stable in Pvt FT, limited transitions to Pvt PT, moderate income mobility (Q5: 18.2\%) \\
            \hline
            \textbf{8}  & $2$ & $4$ & 4.1\% & 62.6\% & 36.2\% over 45 & Low & Limited (Q5: 19.4\%) & High non-employment, women more likely to shift to Pvt PT, limited mobility (Q5: 19.4\%) \\
            \hline
            \textbf{9} & $3$ & $1$  & 24.3\% & 64.4\% & 77.2\% under 45 & Mixed & Upward (Q5: 25.5\%) & Diverse transitions, women shift to/from public roles, upward income mobility (Q5: 25.5\%) \\
            \hline
            \textbf{10}  & $3$ & $2$  & 2.5\% & 65.0\% & 80.2\% under 45 & Low-Mid & Limited (Q5: 13.1\%) & High non-employment, younger workforce, women less upwardly mobile (Q5: 13.1\%) \\
            \hline
            \textbf{11}  & $3$ & $3$  & 32.2\% & 32.4\% & 77.6\% under 45 & Low & Moderate (Q5: 14.4\%) & Men stable in Pvt FT, fewer transitions to public roles, moderate mobility (Q5: 14.4\%) \\
            \hline
            \textbf{12}  & $3$ & $4$ & 1.1\% & 55.9\% & 83.3\% under 45 & Mixed & Moderate (Q5: 20.1\%) & Diverse transitions, balanced gender, younger workforce, moderate upward mobility (Q5: 20.1\%) \\
            \hline
            \end{tabular}
            \label{tab:latent_class_summary}
        \end{table}

        \begin{table}[ht!] 
    \centering
    \scriptsize
    \setlength{\tabcolsep}{4pt} 
    \renewcommand{\arraystretch}{1.1} 
    \caption{Predicted Transition Probabilities (Aggregate, Men, Women)}
    \begin{tabular}{@{}lccccc@{}}
    \toprule
    \textbf{From/To}     & NE & Pvt FT & Pub FT & Pvt PT & Pub PT \\
    \midrule
    \multicolumn{6}{c}{\textbf{Aggregate}} \\
    \midrule
    Non-employment (NE)       & \centering 0.82  & 0.10       & 0.02      & 0.04      & 0.01      \\
    Private full-time (Pvt FT) & \centering 0.08  & 0.89       & 0.0025     & 0.03      & 0.0005    \\
    Public full-time (Pub FT)  & \centering 0.02  & 0.01       & 0.93      & 0.0015     & 0.04      \\
    Private part-time (Pvt PT) & \centering 0.17  & 0.13       & 0.005     & 0.69       & 0.005      \\
    Public part-time (Pub PT)  & \centering 0.09  & 0.01       & 0.14      & 0.01      & 0.74      \\
    \textit{Total} & \centering \textit{0.31 }& \textit{0.41} & \textit{0.14} & \textit{0.10} &\textit{ 0.04} \\
    \bottomrule
    \end{tabular}
    \vspace{1em} 
    \begin{tabular}{@{}lccccc|ccccc@{}}
    \toprule
    & \multicolumn{5}{c}{\textbf{Men}} & \multicolumn{5}{c}{\textbf{Women}} \\
    \cmidrule(lr){2-6} \cmidrule(lr){7-11}
    \textbf{From/To} & NE & Pvt FT & Pub FT & Pvt PT & Pub PT & NE & Pvt FT & Pub FT & Pvt PT & Pub PT \\
    \midrule
    NE       & \centering 0.84 & 0.11 & 0.015 & 0.03 & 0.005 & 0.82 & 0.09 & 0.02 & 0.06 & 0.02 \\
    Pvt FT   & \centering 0.07 & 0.91  & 0.002 & \textcolor{red}{\textbf{0.02}} & 0.0003 & 0.09 & 0.86 & 0.003 & \textcolor{red}{\textbf{0.05}} & 0.001 \\
    Pub FT   & \centering 0.03 & 0.007 & 0.94 & 0.001 & \textcolor{red}{\textbf{0.02}} & 0.02 & 0.005 & 0.93 & 0.002 & \textcolor{red}{\textbf{0.05}} \\
    Pvt PT   & \centering 0.24 & \textcolor{blue}{\textbf{0.21}} & 0.006 & \textbf{0.54} & 0.003 & 0.15 & \textcolor{blue}{\textbf{0.10}} & 0.005 & \textbf{0.74} & 0.01 \\
    Pub PT   & \centering 0.17 & 0.02 & 0.20 & 0.01 & \textbf{0.60} & 0.08 & 0.01 & 0.13 & 0.01 & \textbf{0.77} \\
    \textit{Total} & \centering \textit{0.32 }& \textit{0.52} & \textit{0.11} & \textit{0.04} &\textit{ 0.01}& \textit{0.31} & \textit{0.31} & \textit{0.17} &\textit{ 0.15} & \textit{ 0.06}\\
    \bottomrule
     \multicolumn{10}{c}{\textbf{Max distance from observed: Aggregate=0.01 , Men=0.07 , Women=0.01} }
    \end{tabular}
    \label{tab_trans_proba_pred}
\end{table}

        \begin{figure}[H]
        \centering
        \includegraphics[width=.7\linewidth]{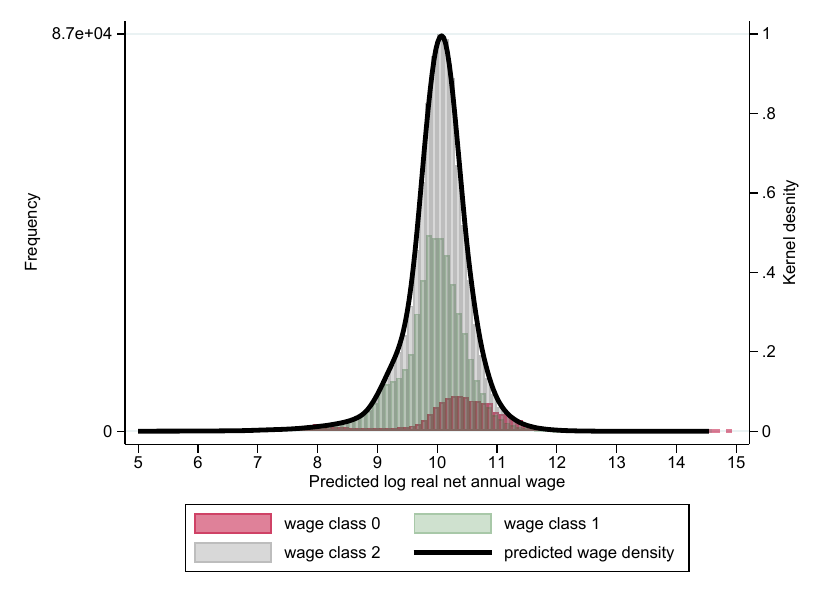}
        \caption{Predicted latent wage class distribution}
        \label{fig:pred_wage_latent}
    \end{figure}

    \begin{figure}[H]
    \centering
    \begin{subfigure}[b]{0.45\textwidth}
        \centering
        \includegraphics[width=\textwidth]{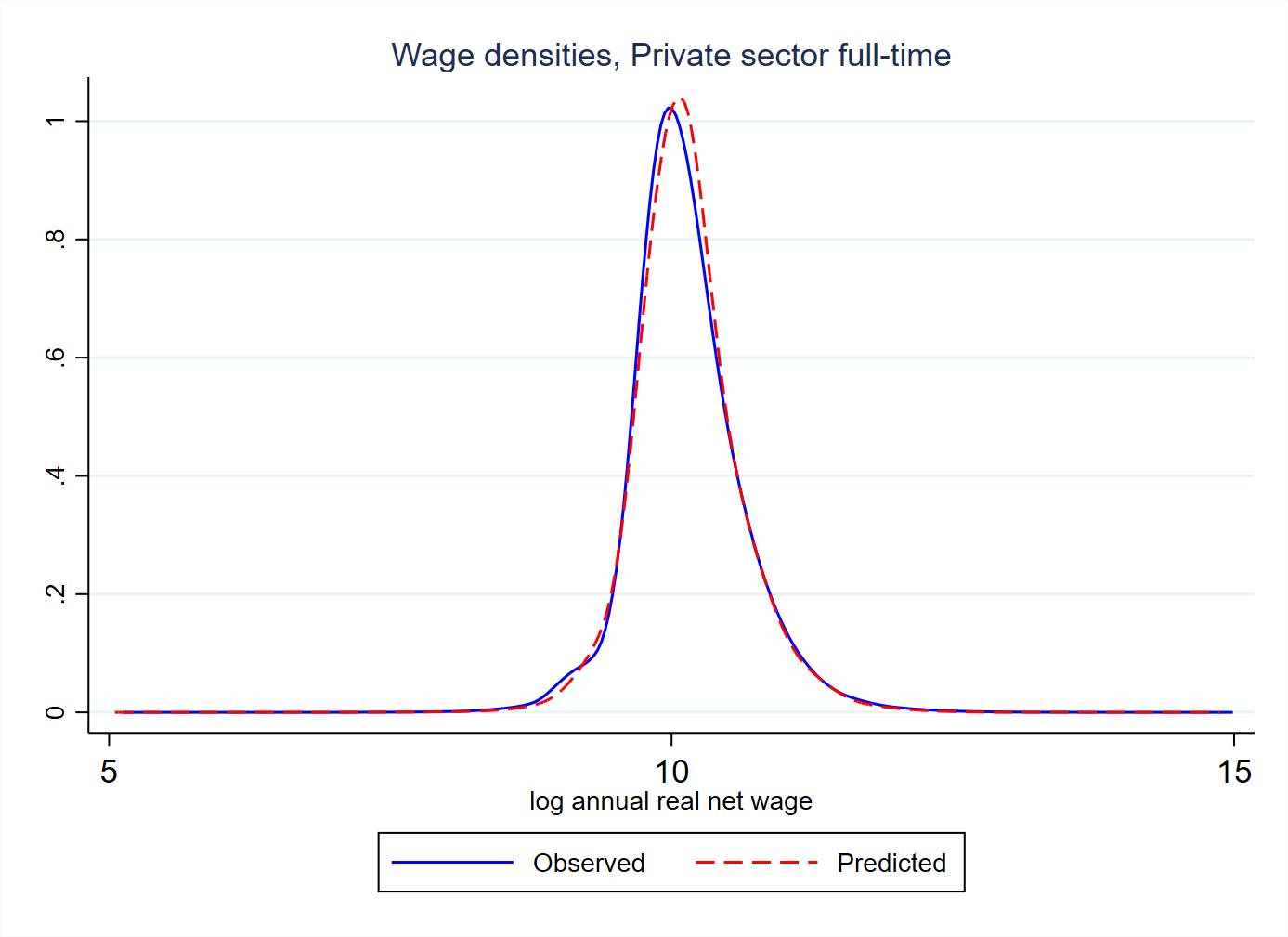}
    \end{subfigure}
    \begin{subfigure}[b]{0.45\textwidth}
        \centering
        \includegraphics[width=\textwidth]{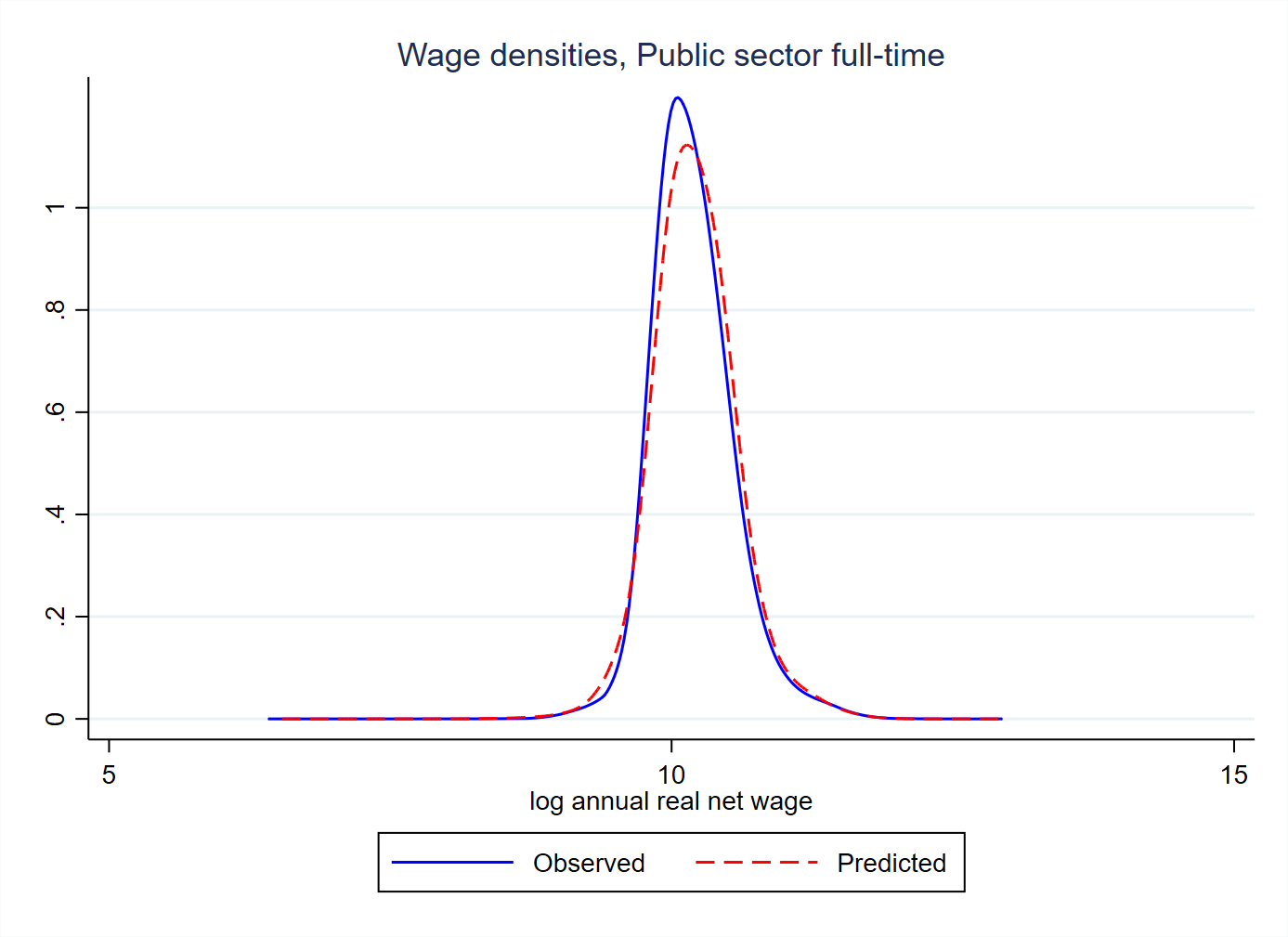}
    \end{subfigure}    
    \begin{subfigure}[b]{0.45\textwidth}
        \centering
        \includegraphics[width=\textwidth]{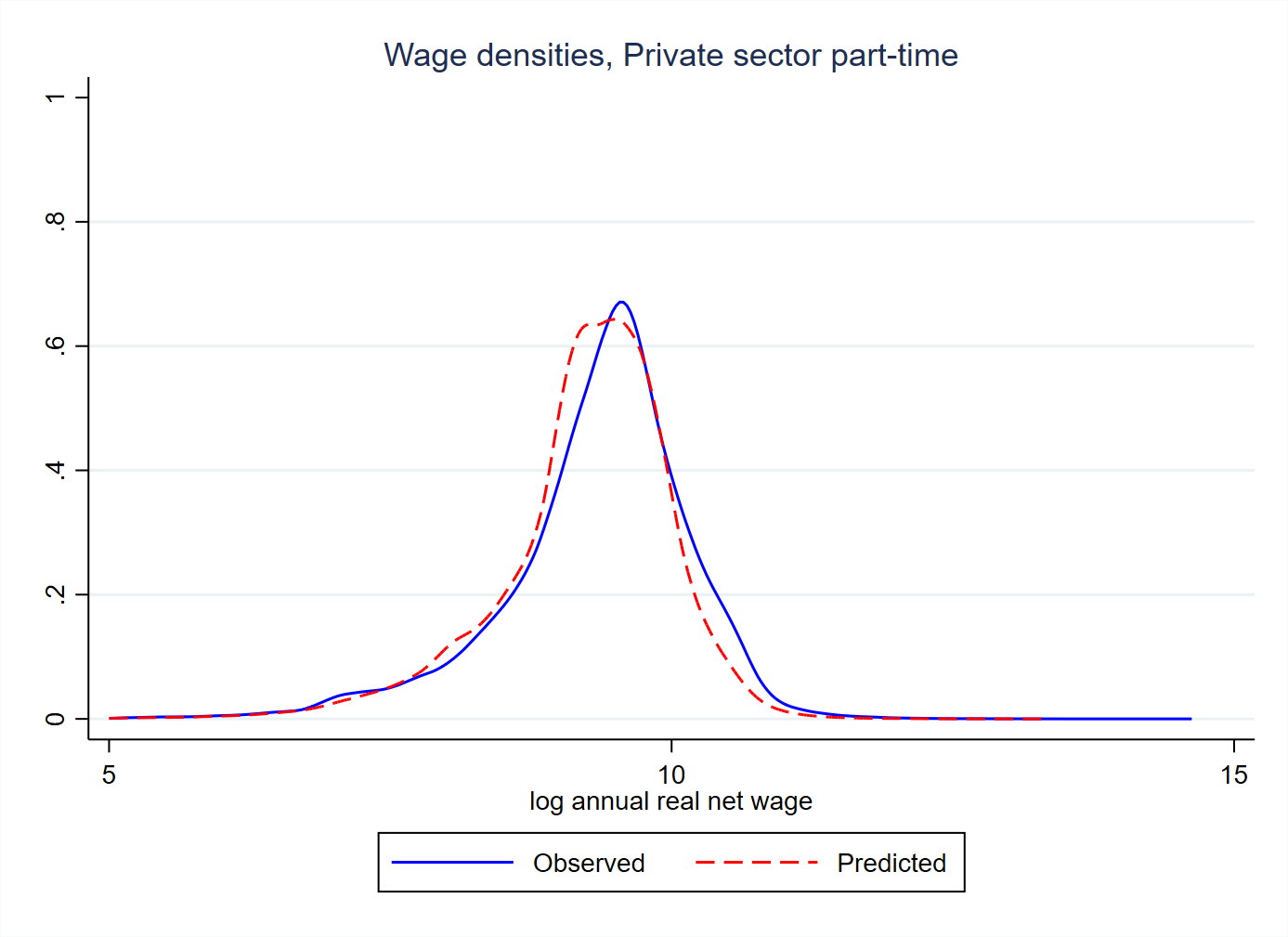}
    \end{subfigure}
    \begin{subfigure}[b]{0.45\textwidth}
        \centering
        \includegraphics[width=\textwidth]{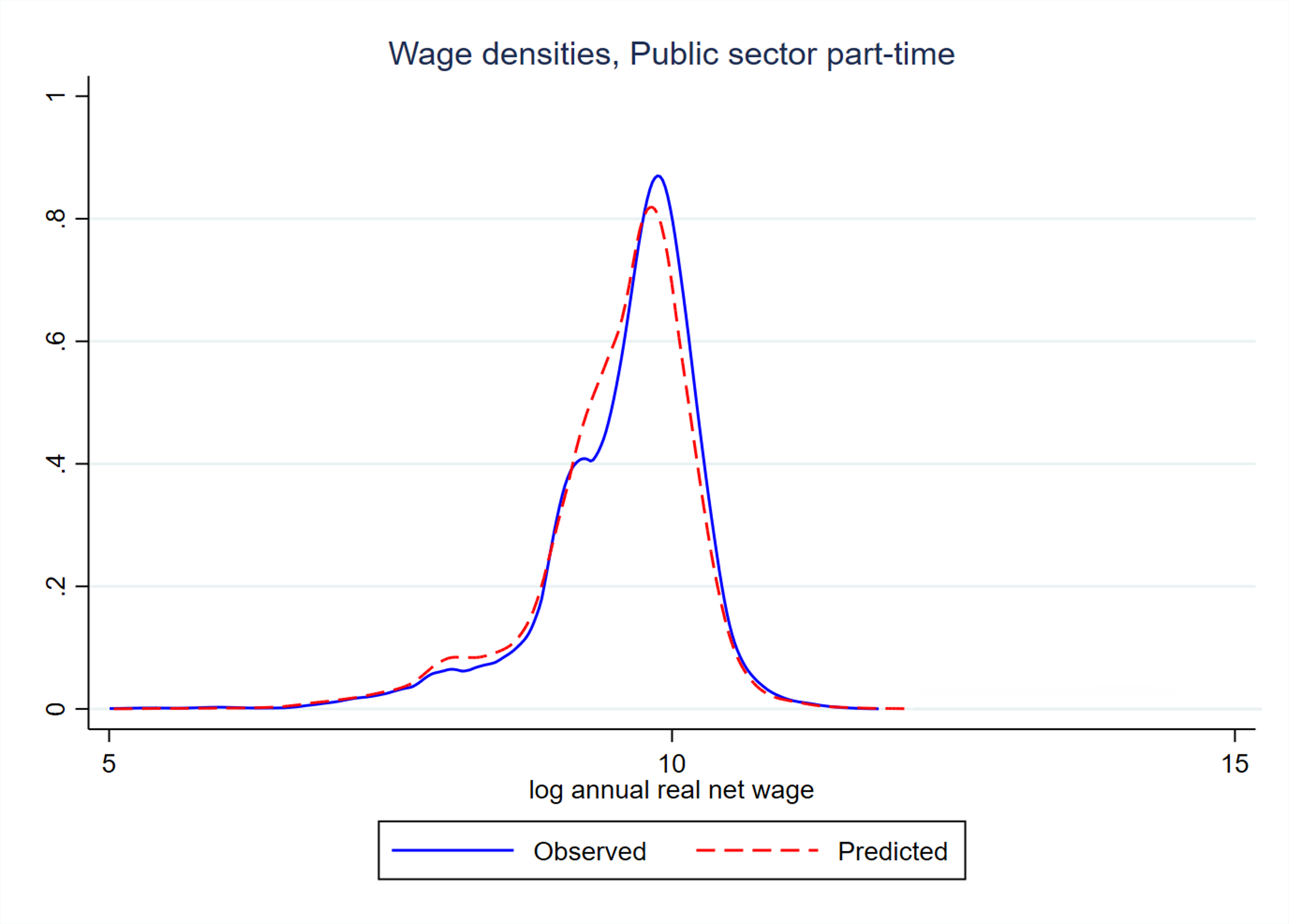}
    \end{subfigure} 
    \caption{Wage Densities by employment state: Observed vs Predicted}
    \label{fig:density_by_state_obs_pred}
\end{figure}

    \subsubsection{Prediction: Out-of-sample predictions of wages and transitions} \label{sec_oos}
     \textbf{Transition probabilities}
            \begin{table}[H] 
    \centering
    \scriptsize
    \setlength{\tabcolsep}{4pt}
    \renewcommand{\arraystretch}{1.1}
    \caption{Transition Probabilities (Aggregate, Men, Women)}
    \begin{tabular}{@{}lccccc@{}}
    \toprule
    \textbf{From/To}     & NE & Pvt FT & Pub FT & Pvt PT & Pub PT \\
    \midrule
    \multicolumn{6}{c}{\textbf{Aggregate}} \\
    \midrule
    Non-employment (NE)       & 0.83 & 0.10 & 0.02 & 0.04 & 0.01 \\
    Private full-time (Pvt FT) & 0.08 & 0.89 & 0.0024 & 0.03 & 0.0005 \\
    Public full-time (Pub FT)  & 0.02 & 0.01 & 0.93 & 0.0014 & 0.04 \\
    Private part-time (Pvt PT) & 0.17 & 0.13 & 0.01 & 0.69 & 0.01 \\
    Public part-time (Pub PT)  & 0.09 & 0.01 & 0.14 & 0.01 & 0.74 \\
    \textit{Total} & \textit{0.31} & \textit{0.41} & \textit{0.14} & \textit{0.10} & \textit{0.04} \\
    \bottomrule
    \end{tabular}
    \vspace{1em}
    \begin{tabular}{@{}lccccc|ccccc@{}}
    \toprule
    & \multicolumn{5}{c}{\textbf{Men}} & \multicolumn{5}{c}{\textbf{Women}} \\
    \cmidrule(lr){2-6} \cmidrule(lr){7-11}
    \textbf{From/To} & NE & Pvt FT & Pub FT & Pvt PT & Pub PT & NE & Pvt FT & Pub FT & Pvt PT & Pub PT \\
    \midrule
    NE       & 0.84 & 0.11 & 0.02 & 0.03 & 0.01 & 0.82 & 0.09 & 0.02 & 0.06 & 0.02 \\
    Pvt FT   & 0.07 & 0.91 & 0.0019 & 0.02 & 0.0002 & 0.09 & 0.86 & 0.0034 & 0.05 & 0.0009 \\
    Pub FT   & 0.03 & 0.01 & 0.94 & 0.0011 & 0.02 & 0.02 & 0.0045 & 0.93 & 0.0015 & 0.05 \\
    Pvt PT   & 0.24 & 0.22 & 0.01 & 0.54 & 0.0033 & 0.14 & 0.10 & 0.0049 & 0.74 & 0.01 \\
    Pub PT   & 0.17 & 0.02 & 0.20 & 0.01 & 0.59 & 0.08 & 0.01 & 0.13 & 0.01 & 0.77 \\
    \textit{Total} & \textit{0.32} & \textit{0.52} & \textit{0.11} & \textit{0.04} & \textit{0.01} & \textit{0.31} & \textit{0.31} & \textit{0.17} & \textit{0.15} & \textit{0.06} \\
    \bottomrule
     \multicolumn{10}{c}{\textbf{Max distance from observed: Aggregate=0.01 , Men=0.08 , Women=0.01}}
    \end{tabular}
    \label{tab_trans_proba_pred_OS}
\end{table}

        \textbf{Wage prediction fit}

        \begin{figure}[H]
            \centering
            \includegraphics[width=0.8\linewidth]{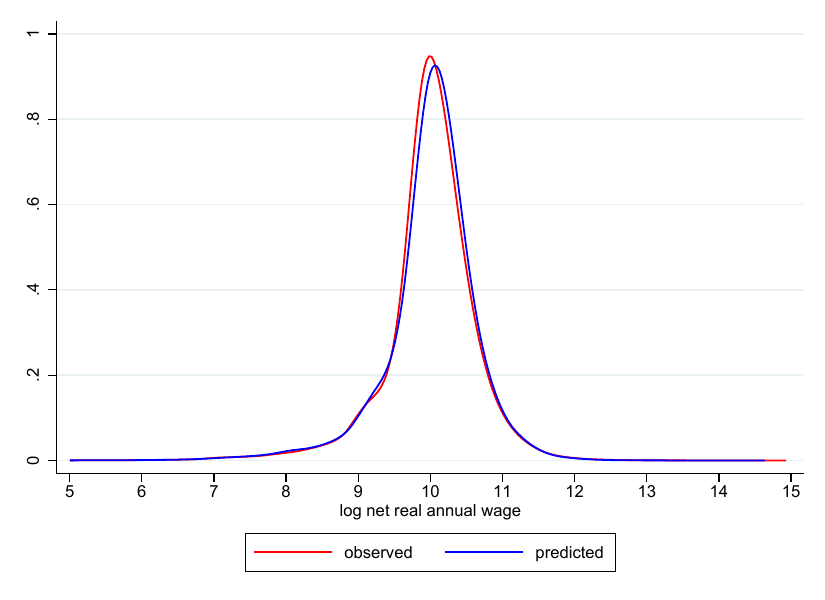}
            \caption{Observed and predicted wage densities}
            \label{fig:ob_pred_dens}
        \end{figure}
         \begin{figure}[H]
            \centering
            \includegraphics[width=0.8\linewidth]{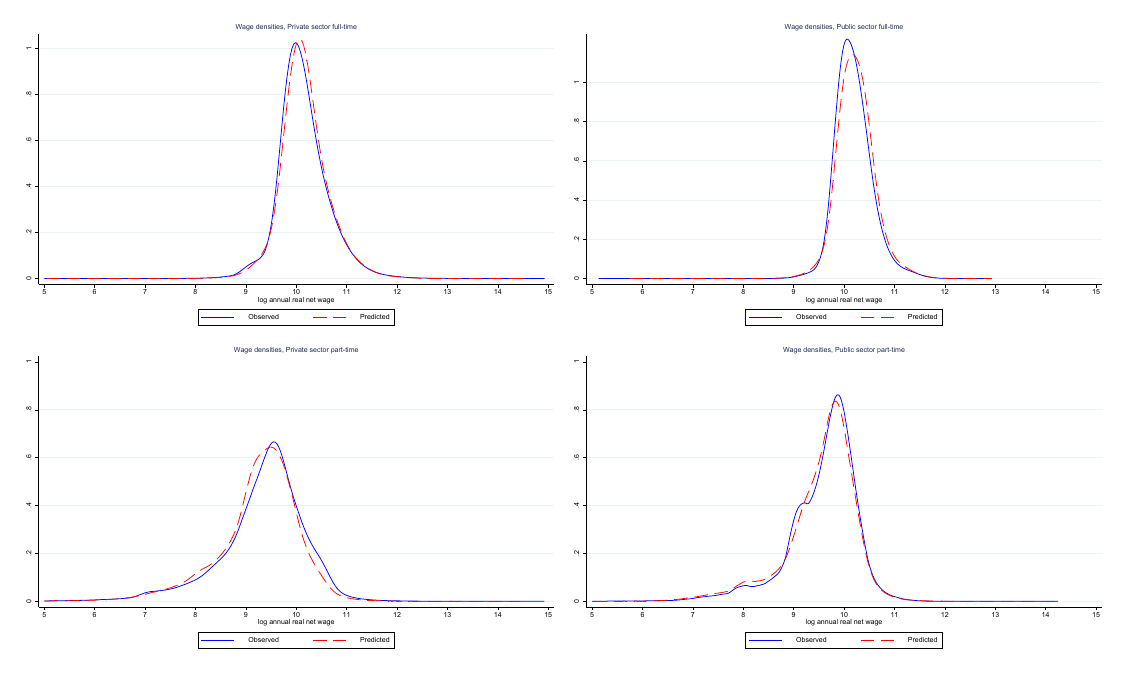}
            \caption{Observed and predicted wage densities by employment state}
            \label{fig:ob_pred_dens_emp}
        \end{figure}
         \begin{figure}[H]
            \centering
            \includegraphics[width=0.8\linewidth]{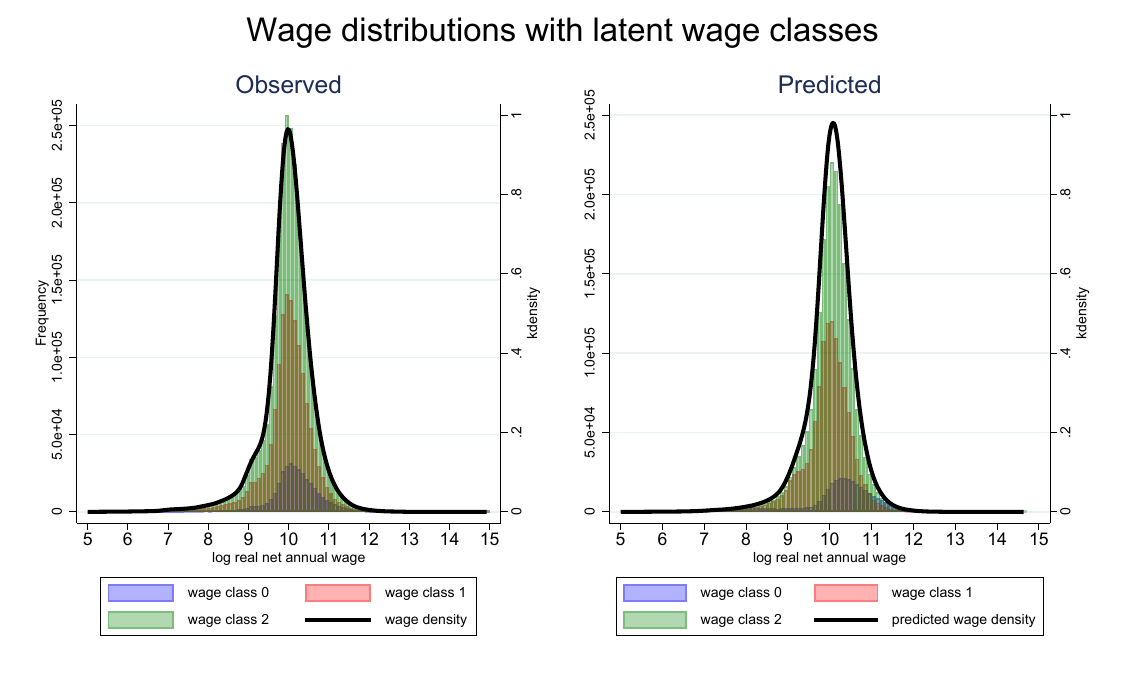}
            \caption{Observed and predicted wage densities and wage classes}
            \label{fig:ob_pred_dens_wage_class}
        \end{figure}
    \subsection{Estimated model parameters}
        \begin{table}[htbp]\centering
        \footnotesize
        \caption{Parameters of unobserved mobility heterogeneity (mlogit models)}
        \begin{tabular}{llll}
        \hline
        \hline
        \multicolumn{4}{l}{Mobility heterogeneity: $\kappa_i^m | z_i^f$}  \\ \hline \hline
        \multicolumn{4}{l}{$\kappa_i^m=1$}                                      \\ \hline
        female               & -0.233            & first xp              & -0.267              \\
        educ=1            & 0.198             & constant            & -1.361            \\
        educ=2            & -20.756             &                     &                    \\ \hline
        \multicolumn{4}{l}{$\kappa_i^m=2$}                                       \\ \hline
        female               & -1.581             & first xp              & -0.263             \\
        educ=1            & -0.626             & constant            & 1.856          \\
        educ=2            & -20.980            &                     &                    \\ \hline
        \multicolumn{4}{l}{$\kappa_i^m=3$}                                       \\ \hline
        female               & -0.458              & first xp              & 0.060            \\
        educ=1            & -0.758             & constant            & -1.307            \\
        educ=2            & -0.679             &                     &                    \\ \hline
        \end{tabular}
        \label{tab:km-table}
        \end{table}
        
        \begin{table}[htbp]\centering
        \footnotesize
        \caption{Parameters of unobserved income heterogeneity (mlogit models)}
        \begin{tabular}{llll}
        \hline
        \hline
        \multicolumn{4}{l}{Income heterogeneity: $\kappa_i^y | \kappa_i^m,z_i^f$} \\ \hline \hline
        \multicolumn{4}{l}{$\kappa_i^y=1$}                                                                    \\ \hline
        female                          & 0.752               & km=1                   & 1.438                 \\
        educ=1                       & -0.203               & km=2                   & 0.535                 \\
        educ=2                       & -0.968                & km=3                   & -0.628                 \\
        first xp                       & -0.354              & constant               & 1.835                \\ \hline
        $\kappa_i^y=2$               &                       &                        &                       \\ \hline
        female                          & 0.535                & km=1                   & 1.310                 \\
        educ=1                       & -0.372               & km=2                   & 0.984                \\
        educ=2                       & -0.497                 & km=3                   & -2.588                 \\
        first xp                       & -0.902              & constant               & 3.312
                \\ \hline       
        \end{tabular}
        \label{tab:ky-table}
        \end{table}
        
        \begin{table}[htbp]\centering
        \footnotesize
        \centering
        \caption{Parameters of state mobility (mlogit models) }
        \label{tab:chi-table}
        \resizebox{\textwidth}{!}{
        \begin{tabular}{llll|llll}
        \hline
        \hline
        \multicolumn{8}{l}{Initial state selection: Pr\{$S_{i1}$\}}                                \\ \hline
        \hline
        \multicolumn{4}{l|}{$S_{i1}=1$}               & \multicolumn{4}{l}{$S_{i1}=3$}               \\ \hline
        female           & -0.609  & $k^m$=1  & -0.147  & female           & 0.958   & $k^m$=1  & -0.203  \\
        med educ         & 0.653   & $k^m$=2  & 0.108   & med educ         & 0.170   & $k^m$=2  & -0.197  \\
        high educ        & 0.695   & $k^m$=3  & -0.917  & high educ        & 0.068   & $k^m$=3  & 0.596   \\
        first xp         & 0.600   & constant & -0.281  & first xp         & 0.480   & constant & -1.912  \\ \hline
        \multicolumn{4}{l|}{$S_{i1}=2$}               & \multicolumn{4}{l}{$S_{i1}=4$}               \\ \hline
        female           & 0.233   & $k^m$=1  & -0.059  & female           & 1.298   & $k^m$=1  & 0.004   \\
        med educ         & 1.175   & $k^m$=2  & -0.259  & med educ         & 0.866   & $k^m$=2  & -0.381  \\
        high educ        & 1.635   & $k^m$=3  & -2.324  & high educ        & 0.925   & $k^m$=3  & -0.440  \\
        first xp         & 0.958   & constant & -2.676  & first xp         & 0.608   & constant & -3.715  \\ \hline
        \hline
        \multicolumn{8}{l}{State selection parameters: Pr\{$S_{it}$\} t\textgreater{}1}            \\ \hline
        \hline
        \multicolumn{4}{l|}{$S_{it}=1$}               & \multicolumn{4}{l}{$S_{it}=3$}               \\ \hline
        $S_{i,t-1}$=1    & 3.119   & $xp^2$    & -0.119  & $S_{i,t-1}$=1    & 1.379   & $xp^2$    & -0.059  \\
        $S_{i,t-1}$=2    & 0.231   & female    & -0.264  & $S_{i,t-1}$=2    & -0.587  & female    & 0.731   \\
        $S_{i,t-1}$=3    & 1.391   & med educ  & 0.370   & $S_{i,t-1}$=3    & 3.282   & med educ  & 0.175   \\
        $S_{i,t-1}$=4    & -0.171  & high educ & 0.439   & $S_{i,t-1}$=4    & 0.368   & high educ & 0.134   \\
        $xp_{i,t-1}$     & 1.675   & first xp  & -1.545  & $xp_{i,t-1}$     & 0.984   & first xp  & -0.819  \\
        $xp_{i,t-1}*S_{i,t-1}$=1 & 0.704 & $k^m$=1 & -0.110 & $xp_{i,t-1}*S_{i,t-1}$=1 & 0.298 & $k^m$=1 & -0.110 \\
        $xp_{i,t-1}*S_{i,t-1}$=2 & 0.081 & $k^m$=2 & 0.024  & $xp_{i,t-1}*S_{i,t-1}$=2 & 0.353 & $k^m$=2 & -0.144 \\
        $xp_{i,t-1}*S_{i,t-1}$=3 & 0.235 & $k^m$=3 & -0.601 & $xp_{i,t-1}*S_{i,t-1}$=3 & 0.485 & $k^m$=3 & 0.183  \\
        $xp_{i,t-1}*S_{i,t-1}$=4 & -0.242 & constant & -2.065 & $xp_{i,t-1}*S_{i,t-1}$=4 & 0.104 & constant & -3.506 \\ \hline
        \multicolumn{4}{l|}{$S_{it}=2$}               & \multicolumn{4}{l}{$S_{it}=4$}              \\ \hline
        $S_{i,t-1}$=1    & -0.270  & $xp^2$    & -0.170  & $S_{i,t-1}$=1    & -1.568  & $xp^2$    & -0.124  \\
        $S_{i,t-1}$=2    & 5.307   & female    & 0.261   & $S_{i,t-1}$=2    & 3.686   & female    & 0.947   \\
        $S_{i,t-1}$=3    & 0.237   & med educ  & 0.494   & $S_{i,t-1}$=3    & 0.363   & med educ  & 0.462   \\
        $S_{i,t-1}$=4    & 2.930   & high educ & 0.769   & $S_{i,t-1}$=4    & 4.669   & high educ & 0.447   \\
        $xp_{i,t-1}$     & 1.063   & first xp  & -0.906  & $xp_{i,t-1}$     & 1.254   & first xp  & -0.936  \\
        $xp_{i,t-1}*S_{i,t-1}$=1 & 0.333 & $k^m$=1 & -0.023 & $xp_{i,t-1}*S_{i,t-1}$=1 & 0.381 & $k^m$=1 & 0.018  \\
        $xp_{i,t-1}*S_{i,t-1}$=2 & 1.229 & $k^m$=2 & -0.130 & $xp_{i,t-1}*S_{i,t-1}$=2 & 0.514 & $k^m$=2 & -0.296 \\
        $xp_{i,t-1}*S_{i,t-1}$=3 & -0.067 & $k^m$=3 & -1.058 & $xp_{i,t-1}*S_{i,t-1}$=3 & 0.154 & $k^m$=3 & -0.125 \\
        $xp_{i,t-1}*S_{i,t-1}$=4 & 0.745 & constant & -4.010 & $xp_{i,t-1}*S_{i,t-1}$=4 & 0.782 & constant & -5.101 \\ \hline
        \end{tabular}
        }
        \end{table}

       \begin{table}[htbp]
        \footnotesize
        \centering
        \caption{Parameters of cross-sectional income means and standard deviations}
        \label{tab:musigma-table}
        \resizebox{\textwidth}{!}{
        \begin{tabular}{llllllll}
        \hline
        \hline
        \multicolumn{8}{l}{Income means: $\mu$}                                                                                                                     \\ \hline \hline
        $S_{it}=1$     & \multicolumn{1}{l|}{0}       & $xp$           & \multicolumn{1}{l|}{0.631}   & \multicolumn{4}{l}{$S_{it} \times k^y$}                         \\
        $S_{it}=2$     & \multicolumn{1}{l|}{-0.079}  & $xp*S_{it}=2$  & \multicolumn{1}{l|}{-0.055}  & \textbf{1 1} & \multicolumn{1}{l|}{-1.061} & \textbf{3 1} & 1.075  \\
        $S_{it}=3$     & \multicolumn{1}{l|}{-3.031}  & $xp*S_{it}=3$  & \multicolumn{1}{l|}{0.096}   & \textbf{1 2} & \multicolumn{1}{l|}{-0.636} & \textbf{3 2} & 1.841  \\
        $S_{it}=4$     & \multicolumn{1}{l|}{-2.683}  & $xp*S_{it}=4$  & \multicolumn{1}{l|}{-0.018}  & \textbf{2 1} & \multicolumn{1}{l|}{-0.416} & \textbf{4 1} & 1.828  \\
        $xp^2$         & \multicolumn{1}{l|}{-0.080}  & female         & \multicolumn{1}{l|}{-0.139}  & \textbf{2 2} & \multicolumn{1}{l|}{-0.711} & \textbf{4 2} & 1.281  \\
        med educ       & \multicolumn{1}{l|}{0.283}   & high educ      & \multicolumn{1}{l|}{0.524}   &              & \multicolumn{1}{l|}{}       &              &        \\
        first xp       & \multicolumn{1}{l|}{-0.164}  & constant       & \multicolumn{1}{l|}{10.182}  &              & \multicolumn{1}{l|}{}       &              &        \\ \hline \hline
        \multicolumn{8}{l}{Income standard deviation: $\sigma$}                                                                                                            \\ \hline \hline
        $S_{it}=2$     & \multicolumn{1}{l|}{0.085}   & $xp^2*S_{it}=1$ & \multicolumn{1}{l|}{0.076}  & \multicolumn{4}{l}{$S_{it} \times k^y$}                         \\
        $S_{it}=3$     & \multicolumn{1}{l|}{1.842}   & $xp^2*S_{it}=2$ & \multicolumn{1}{l|}{0.187}  & \textbf{1 1} & \multicolumn{1}{l|}{-0.348} & \textbf{3 1} & -0.582 \\
        $S_{it}=4$     & \multicolumn{1}{l|}{2.082}   & $xp^2*S_{it}=3$ & \multicolumn{1}{l|}{0.122}  & \textbf{1 2} & \multicolumn{1}{l|}{-0.224} & \textbf{3 2} & -0.632 \\
        $xp$           & \multicolumn{1}{l|}{-0.590}  & $xp^2*S_{it}=4$ & \multicolumn{1}{l|}{0.259}  & \textbf{2 1} & \multicolumn{1}{l|}{-0.432} & \textbf{4 1} & -1.552 \\
        $xp*S_{it}=2$  & \multicolumn{1}{l|}{-0.538}  & first xp         & \multicolumn{1}{l|}{0.232}   & \textbf{2 2} & \multicolumn{1}{l|}{-0.341} & \textbf{4 2} & -0.686 \\
        $xp*S_{it}=3$  & \multicolumn{1}{l|}{-0.327}  & $k^m=1$        & \multicolumn{1}{l|}{-0.006}  &              & \multicolumn{1}{l|}{}       &              &        \\
        $xp*S_{it}=4$  & \multicolumn{1}{l|}{-0.788}  & $k^m=2$        & \multicolumn{1}{l|}{-0.054}  &              & \multicolumn{1}{l|}{}       &              &        \\
        $k^m=3$        & \multicolumn{1}{l|}{1.879}   & female         & \multicolumn{1}{l|}{-0.019}  &              & \multicolumn{1}{l|}{}       &              &        \\
        med educ       & \multicolumn{1}{l|}{-0.021}  & high educ      & \multicolumn{1}{l|}{0.178}   &              & \multicolumn{1}{l|}{}       &              &        \\
        constant       & \multicolumn{1}{l|}{-3.524}  &                & \multicolumn{1}{l|}{}        &              & \multicolumn{1}{l|}{}       &              &        \\ \hline \hline
        \end{tabular}
        }
        \end{table}

       \begin{table}
        \footnotesize
        \centering
        \caption{Parameters of income mobility }
        \label{tab:tau-table}
        \resizebox{\textwidth}{!}{
        \begin{tabular}{llllllll}
        \hline \hline
        \multicolumn{8}{l}{First-order income autocorrelation: $\tau_1$}                                                                                                                    \\ \hline \hline
        $k^y=1$     & \multicolumn{1}{l|}{-0.139} & $S_{i,t-1}=1$ & \multicolumn{1}{l|}{0.151}  & \multicolumn{2}{l|}{$k^y \times S_{i,t}$} & \multicolumn{2}{l}{$xp_{i,t-1} \times S_{i,t-1}$} \\
        $k^y=2$     & \multicolumn{1}{l|}{0.008}  & $S_{i,t-1}=2$ & \multicolumn{1}{l|}{0.131}  & \textbf{1 2}  & \multicolumn{1}{l|}{0.077}  & \textbf{1}                & -0.019             \\
        $S_{i,t}=2$ & \multicolumn{1}{l|}{0.019}  & $S_{i,t-1}=3$ & \multicolumn{1}{l|}{0.149}  & \textbf{1 3}  & \multicolumn{1}{l|}{0.126}  & \textbf{2}                & -0.057             \\
        $S_{i,t}=3$ & \multicolumn{1}{l|}{-0.066} & $S_{i,t-1}=4$ & \multicolumn{1}{l|}{0.096}  & \textbf{1 4}  & \multicolumn{1}{l|}{-0.110} & \textbf{3}                & 0.039              \\
        $S_{i,t}=4$ & \multicolumn{1}{l|}{0.204}  & $xp_{i,t-1}$  & \multicolumn{1}{l|}{-0.610} & \textbf{2 2}  & \multicolumn{1}{l|}{-0.069} & \textbf{4}                & 0.008              \\
        $xp$        & \multicolumn{1}{l|}{0.717}  & $k^m=1$       & \multicolumn{1}{l|}{-0.033} & \textbf{2 3}  & \multicolumn{1}{l|}{0.068}  & \multicolumn{2}{l}{}       \\
        $xp^2$      & \multicolumn{1}{l|}{-0.017} & $k^m=2$       & \multicolumn{1}{l|}{-0.035} & \textbf{2 4}  & \multicolumn{1}{l|}{-0.186} & \multicolumn{2}{l}{}       \\
        constant    & \multicolumn{1}{l|}{0.310}  & $k^m=3$       & \multicolumn{1}{l|}{0.194}  & \multicolumn{2}{l|}{}                      & \multicolumn{2}{l}{}       \\ \hline \hline
        \end{tabular}
        }
        \end{table}

\subsection{Results: Figures}
\begin{figure}[H]
    \centering
    \includegraphics[width=\linewidth]{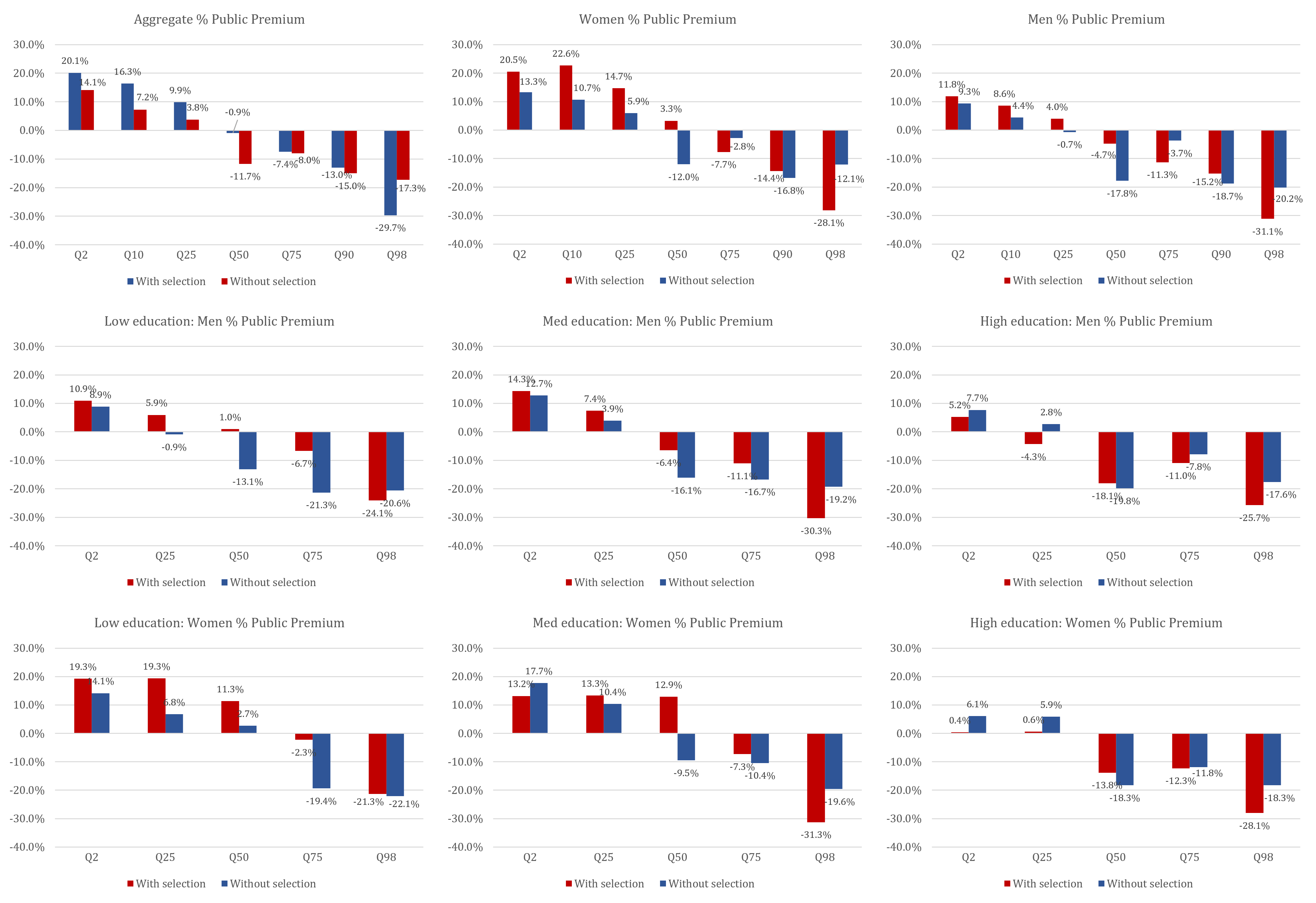}
    \caption{Public premium in lifetime earnings: "job for life" counterfactuals with and without selection}
    \label{fig:w_wo_selection_premialabel}
\end{figure}
\end{document}